\documentclass[fleqn,usenatbib]{mnras}

\usepackage{newtxtext,newtxmath}

\usepackage[T1]{fontenc}

\DeclareRobustCommand{\VAN}[3]{#2}
\let\VANthebibliography\thebibliography
\def\thebibliography{\DeclareRobustCommand{\VAN}[3]{##3}\VANthebibliography}

\usepackage{graphicx}	
\usepackage{amsmath}

\newcommand{\cm}{\,\ifmmode{{\mathrm{cm}}}\else cm\fi}

\newcommand{\ergps}{\,{\rm erg}\,{\rm s}\ifmmode{}^{-1}\else${}^{-1}$\fi}
\newcommand{\Mpch}{\,{\rm Mpc}\,\ifmmode h^{-1}\else $h^{-1}$\fi}
\newcommand{\snru}{\,\ifmmode{\mathrm{Myr}^{-1}}\else Myr${}^{-1}$\fi}
\newcommand{\Kms}{\,\ifmmode{\mathrm{km}\,\mathrm{s}^{-1}}\else km\,s${}^{-1}$\fi}
\newcommand{\sbu}{\,{\rm erg}\,{\rm s}^{-1}\,{\rm cm}^{-2}\,{\rm arcsec}^{-2}}
\newcommand{\Lya}{Ly$\alpha\,$}
\defcitealias{Fielding_2022}{FB22}

\title[A multiphase wind model with multiple clouds]{Strength in numbers: A multiphase wind model with multiple cloud populations}

\author[C. Nikolis \& M. Gronke]{
C. Nikolis,$^{1,2}$\thanks{E-mail: \text{harrynikolis@gmail.com}} and
M. Gronke$^{2}$
\\
$^{1}$Fakultät für Physik, Ludwig-Maximilians-Universität München, Geschwister-Scholl-Platz 1, D-80539 München, Germany\\
$^{2}$Max-Planck-Institut für Astrophysik, Karl-Schwarzschild-Str. 1, D-85741 Garching bei München, Germany
}

\date{Accepted XXX. Received YYY; in original form ZZZ}

\pubyear{2023}

\begin{document}
\label{firstpage}
\pagerange{\pageref{firstpage}--\pageref{lastpage}}
\maketitle

\begin{abstract}
Galactic outflows have a multiphase nature making them challenging to model analytically. Many previous studies have tried to produce models that come closer to reality. 
In this work, we continue these efforts and describe the interaction of the hot wind fluid with multiple cold cloud populations, with their number density determined by different probability density functions. To do so, we introduced realistic cloud-wind interaction source terms and a time varying cooling area. We find that the model reproduces well results from small-scale hydrodynamic simulations, but exhibits a general destructive behaviour both for a single cloud population as well as multiple ones. We show that including multiple cloud populations can alter the evolution of the wind drastically.
We also compare our model to observations and show that the differential acceleration of multiple clouds can lead to a non-negligible velocity `dispersion' relevant for down-the-barrel studies. Furthermore, we compute the emitted cooling surface brightness and find it generally too faint to explain observed Lyman-$\alpha$ halos.
\end{abstract}




\section{Introduction}
Galaxies are not static systems, but in contrast they are subject to complex physical processes, taking place inside them at different scales, that determine their evolution \citep{galaxies6040114}. A process of particular importance is the competition between gas flowing in and out of the galaxy.

The amount of gas and material inside the galaxy is determined by the exchange of gas between inflows and outflows, thus this process is essential for the fate of baryons in the galaxy \citep{Bell:2003je}, as well as for star formation and for regulating the black hole growth. Outflows have their own importance, mainly because they enrich the Circumgalactic and Intergalactic medium with their material \citep{Tumlinson_2017} and subsequently affect the star formation of their host galaxies as well as because they help us understand the mechanisms that powered them.

The main mechanism powering the outflows is thought to be feedback from star formation \citep{galaxies6040138}, or AGN \citep{doi:10.1146/annurev-astro-082214-122316}. Using different probes, observations show that these outflows exhibit a vast range of temperatures. An example of different probes are neutral hydrogen (e.g \citet{Walter_2002}) for $T\sim 10-10^2$K, photo-ionised metals, for $T\sim 10^4-10^5K$ \citep{1995A&A...293..703M,Martin_2009,Westmoquette_Smith_Gallagher_2011} and X-rays for $T\sim 10^7K$ (e.g \citet{2009ApJ...697.2030S}, for more details see \citet{2005ARA&A..43..769V,Veilleux_2020} and references therein). These observational findings lead us to the conclusion that the galactic winds are essentially multiphase. However the multiphase nature of the winds makes the theoretical modelling of this kind of system a challenging task.

That is because the cool gas is subject to processes that tend to destroy it such as hydrodynamical instabilites (Kelvin-Helmholtz, Rayleigh-Taylor \citet{1961hhs..book.....C}). The destruction process of an initially static cold "cloud" with size $r_{\rm cloud}$ destroyed by mixing with a hot galactic wind of velocity $v_{\rm wind}$ is characterized by a timescale of $t_{\rm cc}\sim \chi^{1/2}r_{\rm cloud}/v_{\rm wind}$, with $\chi$ the density contrast of the two phases, usually taking values of at least $\chi \sim 100$. The characteristic time needed for the hot wind to accelerate the cloud, and become entrained by the wind is given by $t_{\rm drag}\sim \chi r_{\rm cloud}/v_{\rm wind} $, so the destruction time is shorter by a factor of $\chi^{1/2}$ leading to the destruction of the cold phase \citep{1994ApJ...420..213K}.

A new approach to the problem is by including radiative cooling to the process of mixing, which leads to the formation of a radiative turbulent mixing layer. This mixing layer is an intermediate area between phases, with temperature at the geometric mean between the two components $T_{\rm mix}\sim \sqrt{T_hT_c}$ \citep{1990MNRAS.244P..26B}. From the nature of the cooling curves, this intermediate temperature gas cools much faster than the cooling rate of each individual component. In this regime, a parameter space exists in which, mass can be condensed out of the hot phase, causing the cold gas to grow and eventually survive, with a lot of effort being done on characterizing the growth with different survival criteria \citep{2010MNRAS.404.1464M,2016MNRAS.462.4157A,Gronke_2018,Gronke_2019,2020MNRAS.492.1841L,2021MNRAS.501.1143K,2020MNRAS.499.4261S,Abruzzo_2022}.

The presence of turbulence and cooling, together with the complexity of the hydrodynamic equations make the system too difficult to model theoretically. Most of our insights for phenomena taking place during the interactions of the different phases, come from an extensive study of small scale hydrodynamic simulations, with or without cooling (see for example \citealp{Scannapieco_2015,Br_ggen_2016,Goldsmith_2017,2017ApJ...834..144S,Braspenning_2023}
and \citet{doi:10.1146/annurev-astro-081913-040019} and references therein). These simulations study the interactions for a vast set of idealized initial conditions and provide us intuition and dependencies that can be used in most general cases. The most quantitative results in terms of characterizing the mass transfer rates between the phases have come out from simulations focusing on the radiative turbulent mixing layers \citep{Ji_Oh_Masterson_2019,Mandelker,Fielding_2020,Tan_2021}. The results of these simulations have offered us a deeper understanding and precise scalings for the characteristic mixing velocity between the two phases in the RTMLs. Despite the focus on simulations for solving the "cloud-crushing problem", an analytical model describing the multiphase nature is still useful, as it will provide a straightforward way to test predictions and compare with observations.

Various approaches to model theoretically the galactic outflows have been made in the past, at first by using only a hot fluid \citet{Chevalier1985}, and then by using a radiative cooling component \citep{1995ApJ...444..590W, Silich_2003, Silich_2011,Thompson_Quataert_Zhang_Weinberg_2015,Bustard_2016}. Attempts have been made in the past to model the multiphase nature of these outflows (see e.g \citet{Cowie1981, 1996ApJ...463..528S,Zhang_Thompson_Quataert_Murray_2017}. The most recent attempt has been by using a two component prescription, a hot fluid and a cold cloud interacting with each other, including radiative cooling and gravity by \citet{Fielding_2022}  henceforth referred to as \citetalias{Fielding_2022}. 

In \citetalias{Fielding_2022}, the authors have developed the complete set of hydrodynamic equations governing the evolution both of the hot galactic wind, as well as the cold clouds embedded inside it, including terms that account for the interaction between the two components. Despite the number of equations involved in these highly complicated interactions, the evolution of the clouds is characterized by astounding simplicity, as the survival or destruction of the clouds is based solely on a $\xi=\frac{t_{\rm turb}}{t_{\rm cool,mix}}$ factor. The major finding of this model is that heavier clouds can survive and be entrained by the wind, while lighter get destroyed. This is in accordance with the most survival criteria, that connect cloud survival with the initial geometry of the cold clump.

We build upon that, and present a model which consists of a hot fluid, and an arbitrary number of cold clouds, interacting with the hot one. Each cold cloud is modelled as a clump population, with each population having a different initial mass, and the number density of each population is determined by a probability distribution. 

This paper is structured as follows: First we present the analytic model, with details given on the construction of the cloud-wind interaction source terms, an equation regarding the tail evolution, and the introduction of multiple cloud populations. Then we discuss about the results of the model, starting from various limiting cases. In the beginning, we solve the model for a homogeneous background, in order to compare it with hydrodynamic simulations. Then we proceed to a single cloud case, that enables us to understand the phenomena at play and compare it to recent models. After these two limiting cases, we solve the full model for different cloud populations and various distributions. Lastly, we calculate the the emitted luminosity and surface brightness along the line of sight, in order to compare the model with observations. The final part includes a discussion about implications of the model, different physical phenomena that came up and  future directions.

\section{Model}
\label{section: model}
We will build the model step by step, with the goal of introducing a distribution of cloud populations interacting with the wind.

Starting from the initial conditions, we assume a \citet{Chevalier1985} model inside the galactic disc, which will provide the initial values for the wind quantities, when the clouds are introduced.

The clouds are introduced gradually just outside the galactic disc and each population interacts with the wind dynamically. Therefore, in order to describe the system, we need equations for the evolution of both the wind and the clouds.

For the wind perspective, we will use throughout the general steady-state fluid equations including gravity, heating and cooling:
\begin{equation}
\frac{1}{r^2}\frac{\partial }{\partial r}\left(\rho vr^2\right)=\dot{\rho}
\label{mass_cons}
\end{equation}
\begin{equation}
    \frac{1}{r^2}\frac{\partial}{\partial r}\left(r^2\rho v^2\right) +\frac{\partial P}{\partial r}={-\rho \frac{v^2_c}{r}}+\dot{p}
\label{momentum_cons}
\end{equation}
\begin{equation}
    \frac{1}{r^2}\frac{\partial}{\partial r}\left(\rho v r^2 \left(\frac{1}{2}v^2+\frac{\gamma}{\gamma-1}\frac{P}{\rho}-\frac{1}{2}v^2_{\rm esc}\right)\right)=\dot{q}-\mathcal{L}
\label{energy_cons}
\end{equation}
\begin{equation}
    \frac{1}{r^2}\frac{d\left(r^2\rho_{\rm Z} v\right)}{dr}=\dot{\rho}_{\rm Z}=\dot{\rho}_+Z_{\rm cloud}-\dot{\rho}_{-}Z
\end{equation}     
Here, $\dot{\rho},\dot{p},\dot{q}$ are the mass, momentum and energy source terms for the cloud wind interaction, and $Z=\frac{\rho_Z}{\rho}$ and $Z_{\rm cloud}$ are the wind and cloud metallicity, respectively, and $v_{\rm esc}$ is the escape velocity related to the gravitational potential, as introduced in \citetalias{Fielding_2022}. 

In our case these interaction terms are due to the presence of multiple cloud populations, and are obtained from results of small-scale hydrodynamic simulations, as will be seen in the next sections.

\subsection{Inside the Galactic Disc}
For the region $r<R_{\rm disc}$, with no gravity ($v_{\rm esc}=0$) and heating-cooling ($\mathcal{L}=0$), assuming no clouds inside the galaxy and with uniform mass and energy injection ($\dot{q}$, $\dot{\rho}$ related only to the uniform injection and $\dot{p}=0$ because of the absence of clouds), the fluid equations reduce to:
\begin{equation}
\frac{1}{r^2}\frac{\partial }{\partial r}\left(\rho vr^2\right)=\dot{\rho}
\end{equation}

\begin{equation}
\rho v \frac{\partial v}{\partial r}=\frac{\partial P}{\partial r}-\dot{\rho}v  
\end{equation}
\begin{equation}
\frac{1}{r^2}\frac{\partial }{\partial r}\left(\rho v r^2 \left(\frac{1}{2}v^2+\frac{\gamma}{\gamma-1}\frac{P}{\rho}\right)\right)=\dot{q}
\end{equation}
which is the well known \citet{Chevalier1985} model. Solving this model will provide the initial conditions for when the clouds are introduced, in the $r>R_{\rm disc}$ region. 

The initial conditions we use are the ones by \citetalias{Fielding_2022}, thus:
\begin{equation}
    \dot{\rho}=\frac{\dot{M}_{\rm hot}}{4\pi R_{\rm disc}^3}=\eta_{\rm mass,hot}\frac{\rm{SFR}}{4\pi R_{\rm disc}^3}
    \label{eta_M}
\end{equation}
with $\mathrm{SFR}$ the star formation rate, and $\eta_{\rm mass,hot}$ the hot mass loading factor $\eta_{\rm mass,hot}=\dot{M_{\rm hot}}/\rm SFR$. 

A similar term $\eta_{\rm mass,cold}$ can be used when we introduce cloud populations. When we deal with a single cloud population, the $\eta_{\rm mass,cold}$ is the cold mass loading factor of this population. However, in the presence of multiple cloud populations $\eta_{\rm mass,cold}$ will describe the total cold mass flux of the system, which means that it will be equal to the sum of the loading factors of each population, multiplied with an appropriate statistical weight. Mainly, the statistical weight for each cold mass loading factor will be determined by the probability to find the different populations having a specific initial mass flux. 

For the energy source term we can conclude for $r<R_{\rm disc}$ with the same reasoning:
\begin{equation}
    \dot{q}=\frac{\dot{Q}}{4\pi R_{\rm disc}^3}=\eta_{\rm energy}\frac{\dot{Q}_{\rm SN}}{100 M_\odot}\rm{SFR}\frac{1}{4\pi R_{\rm disc}^3}
    \label{eta_E}
\end{equation}
where we used the energy loading factor $\eta_{\rm energy}=\frac{\dot{Q}}{\dot{Q}_{\rm SN}}$ and assumed that all the energy is injected from supernovae that release energy $Q_{\rm SN}=10^{51}\text{erg}$ in the time an amount of a total of $100M_\odot$ is formed.

\subsection{Outside the Galactic Disc}
Outside the galactic disc, we must use the full equations~\ref{mass_cons}-\ref{energy_cons}. The source terms now include the presence of the clouds and to find their form, we must focus on how the wind interacts with a single cloud exchanging mass, momentum and energy.

\subsubsection{The cloud-wind interaction}
From the conservation of mass, we can express the growth of the clouds, as mass flowing from the hot phase to the cold phase. Then, the rate of the mass flowing into the clouds can be written as \citep{Gronke_2019}: 
\begin{equation}
   \Dot{M}_{\rm grow}=\rho_{\rm wind}A_{\rm cool}v_{\rm in} 
   \label{mass_growth_rate}
\end{equation}
With the $A_{\rm cool}$ the surface area of the cloud\footnote{Note that $A_{\rm cool}$ is rather an `effective' cloud surface area; see discussion in \citet{Gronke_2019}.} and $v_{\rm in}$ the inflow velocity. From the formed RTML, a turbulence velocity $v_{\rm turb}$ develops. Using results from small-scale hydrodynamic simulations \citet{Tan_2021} the turbulence $v_{\rm turb}$ has the form:
\begin{equation}
v_{\rm turb}=50\Kms\left(\frac{v_{\rm rel}}{150\Kms}\right)^{0.8}\left(\frac{r_{\rm cloud}}{\rm 100\,pc}\right)^{0.2}\left(\frac{t_{\rm cool,cold}}{\rm 0.03\,Myr}
\right)^{-0.2}
\end{equation}
$t_{\rm cool,cold}$ is defined as the cooling $t_{\rm cool}$ with temperature and metallicity fixed by the cloud values: $t_{\rm cool,cold}=t_{\rm cool}\left(T=T_{\rm cloud},Z=Z_{\rm cloud},P\right)$. For the cooling time, we use the cooling curve from \citet{Wiersma_Schaye_Smith_2009} as in \citetalias{Fielding_2022}. We do not employ explicit heating but use a temperature floor instead mimicking the impact.

From the findings of \citet{Tan_2021}, the inflow velocity changes form, depending on whether the system is in the weak or strong cooling regime. The parameter controlling the two regimes is Damkohler number Da \citep{Kuo_Acharya_2012}, which is the ratio of the outer eddy turnover time to the cooling time $\mathrm{Da}=t_{\rm turb}/t_{\rm cool}$. The two regimes correspond to $\mathrm{Da}<1$ and $\mathrm{Da}>1$ for the weak and strong cooling respectively. Motivated by the scalings of \citet{Tan_2021}, we adopt a turnover point $\mathrm{Da}_{\rm mix}$ with:
\begin{equation}
    \mathrm{Da}_{\rm mix}=\frac{\left(\frac{r_{\rm cloud}}{100\,\rm pc}\right)}{\left(\frac{v_{\rm turb}}{30 \Kms}\right)\left(\frac{t_{\rm cool,cold}}{0.03\,\rm Myr}\right)}=\frac{\mathrm{Da}}{109}
\end{equation}
At the turnover point between the two regimes, the inflow velocity is found to be $v_{\rm in}\sim 16\Kms$ \citep{Tan_2021}. We choose a cloud temperature of $T_{\rm cloud}=2\times 10^4\text{K}$.\footnote{ Observations \citep{1995A&A...293..703M,Westmoquette_2009} trace the cold gas to temperatures $\sim 10^4\text{K}$. However, due to uncertainties in heating processes the exact choice of the cold gas temperature is arbitrary(see, e.g., \citealp{Wiersma_Schaye_Smith_2009, 2011piim.book.....D}). Because of the sharp dropoff in the cooling curve between $T\sim 10^4\text{K}$ and $4\times 10^4\text{K}$, this range of temperatures is most likely and used in previous studies (e.g. \citealp{Tan_2021},\citetalias{Fielding_2022}). One might argue that the choice of $T=2\times10^4\text{K}$ is somewhat higher than most of these studies who employ $10^4\text{K}$ but firstly turbulent heating especially at the base of the winds can raise the cold gas temperatures above this, and more importantly, the exact choice does not play a crucial role for the evolution of the wind since the most relevant cooling time is the minimum cooling time $\sim 4\times10^4\text{K}$ \citep{Tan_2021,Farber2022}.}

In order for the two regimes to smoothly connect at a cloud temperature $T_{\rm cloud}=2\times 10^4\text{K}$, the equations for $v_{\rm in}$ are:
\begin{equation}
    v_{\rm in}=13.39\Kms\mathcal{M}_{\rm turb}^{1/2}\left(\frac{r_{\rm cloud}}{100\text{pc}}\right)^{1/2}\left(\frac{t_{\rm cool,cold}}{0.03\text{Myr}}\right)^{-1/2}
    \label{inflow_weak}
    \end{equation}
for $\rm Da_{mix}<1$ and for $\rm Da_{mix}>1$:
\begin{equation}
    v_{\rm in}=12.24\Kms\mathcal{M}_{\rm turb}^{3/4}\left(\frac{r_{\rm cloud}}{100pc}\right)^{1/4}\left(\frac{t_{\rm cool,cold}}{0.03Myr}\right)^{-1/4}
    \label{inflow_strong}
\end{equation}
with $\mathcal{M}_{\rm turb}=v_{\rm turb}/c_{\rm s,c}$. The turbulent velocity
$v_{\rm turb}$ is saturated at high Mach numbers \citep{https://doi.org/10.48550/arxiv.2205.15336}, so when $\mathcal{M}=\frac{v_{\rm rel}}{c_{\rm s,h}}>1$, $v_{\rm turb}\sim\mathcal{M}^0$, thus for $\mathcal{M}>1$ we have:
\begin{equation}
   v_{\rm turb}=50\Kms\left(\frac{c_{\rm s,h}}{150\text{km/s}}\right)^{0.8}\left(\frac{r_{\rm cloud}}{100\text{pc}}\right)^{0.2}\left(\frac{t_{\rm cool,cold}}{0.03\text{Myr}}\right)^{-0.2} .
   \label{vturb-saturation}
\end{equation}
Because even clouds in the fully entrained state ($v_{\rm rel}\rightarrow 0$) show a non-neglibile, constant $\dot M_{\rm cloud}$ due to `pulsations' of the cloud (see discussions in \citealp{Gronke_2019,abruzzo2022taming} and \citealp{Waters_2019,Gronke_Oh_2020,2022arXiv220900732G} for discussions related to the windtunnel and a static setup, respectively), we also use a minimum value for $v_{\rm in}$ from \citet{Gronke_2019}:
\begin{equation}
    v_{\rm min}=0.2c_{\rm s,c}\left(\frac{r_{\rm cloud}}{t_{\rm cool,cold}c_{\rm s,c}}\right)^{0.25}
    \label{vin-min}.
\end{equation}

For the effective area of the cloud $A_{\rm cool}$, we assume that the length of the cloud expands in the direction of the flow and forms a tail, in accordance to hydro-dynamical simulations. This is where the cooling of the hot mass takes place, leading to cloud growth. We, thus, assume
\begin{equation}
    A_{\rm cool}\sim r_{\rm cloud} L=2\pi r_{\rm cloud}L
\end{equation}
with $L$ the length in the direction of the flow. In order to model the dynamics of the area of the cloud, we use the following qualitative arguments based on `cloud crushing' simulations \citet{2010MNRAS.404.1464M,2016MNRAS.462.4157A,Armillotta_Fraternali_Werk_Prochaska_Marinacci_2017,Gronke_2018}: at early times, part of the cloud retains its spherical shape while the tail has a small width. Hence, in that phase the assumption of a cylindrical geometry is an overestimation of the true area. At later times, it is also important to recall that clouds that are destroyed do not have enough mass to supply the tail formation. Because of the above arguments we assume
\begin{equation}
    L(t)=\mathbf{min}\left(L',\frac{M_{\rm cloud}(t)}{M_{\rm cloud}(t=0)}L'\right).
    \label{minimum_length}
\end{equation}
Here, $L'$ describes the evolution of the tail. For this quantity we assume that as the cloud moves in the hot medium, the tail is formed with a velocity equal to the relative velocity between the cloud and the medium. Thus we can write:
\begin{equation}
    \frac{dL'}{dt}=v_{\rm rel}.
\end{equation}
We can translate this relation to a dependence to the galactocentric radius r, covered by the hot volume filling phase:
\begin{equation}
    \frac{dL'}{dr}=\frac{v_{\rm rel}}{v_{\rm cloud}}.
    \label{expansion-equation-radial}
\end{equation}
This qualitative argument is a straightforward generalisation of the relation for the cooling area by \citetalias{Fielding_2022} in order to promote it to a dynamical variable.

The radius of the cloud can evolve according to a cylinder-like geometry \citep{Huang_2020}:
\begin{equation}
    r_{\rm cloud}(t)=\left(\frac{M_{\rm cloud}}{\pi\rho_{\rm cloud}L(t)}\right)^{1/2}.
    \label{cylinder-radius}
\end{equation}
However, as we mentioned before at earlier times, the tail undergoes a rapid expansion with a small width, while part of the cloud retains its spherical shape. By assuming that the radius evolves according to Eq.~\ref{cylinder-radius}, we underestimate the radius of the cloud for the first few $t_{\rm cc}$ and hence the cooling area. Because of the above, we only allow the radius to evolve adiabatically instead of shrinking:
\begin{equation}
    r'_{\rm cloud}=\textbf{max}\left(\left(\frac{M_{\rm cloud}}{\pi\rho_{\rm cloud}L(t)}\right)^{1/2},\left(\frac{M_{\rm cloud,0}}{4/3\pi\rho_{\rm cloud}}\right)^{1/3}\right)
\end{equation}
where we introduced $M_{\rm cloud,0}\equiv M_{\rm cloud}\left(t=0\right)$.

When the tail is not formed, there is no reason to deviate from spherical geometry while the clouds shrink, therefore
\begin{equation}
    r_{\rm cloud}=\textbf{min}\left(r'_{\rm cloud},\left(\frac{M_{\rm cloud}}{4/3\pi\rho_{\rm cloud}}\right)^{1/3}\right)
\end{equation}

With all the above arguments we have constructed the cloud growth term $\dot{M}_{\rm grow}$, which is related to radiative cooling. The next step is to construct a term that is related to the cloud losing mass. This term will be related to the instabilities that tend to destroy the cloud.

For the mass loss term $\Dot{M}_{\rm loss}$, we use the results of the simulations from \citet{Scannapieco_2015}, who find that clouds actually survive on longer timescales than expected from linear arguments \citep{1994ApJ...420..213K}:
\begin{equation}
    \Dot{M}_{\rm loss}=\frac{M_{\rm cloud}}{t_{\rm life}}=\frac{M_{\rm cloud}}{at_{\rm cc}\sqrt{1+\mathcal{M}_{\rm rel}}}.
    \label{mass_loss}
\end{equation}
The fudge factor $a$ is a free parameter in our model, which we set to $a=2$ to match with hydrodynamic simulations.

The total mass exchange rate is therefore
\begin{equation}
    \dot{M}_{\rm cloud}=\dot{M}_{\rm grow}-\dot{M}_{\rm loss}.
\end{equation}
With the above quantities specified, the evolution of the cloud quantities in terms of the volume covered by the expanding wind are derived by \citetalias{Fielding_2022} and are described by, starting with the mass of the cloud:
\begin{equation}
    \frac{\mathrm d{M}_{\rm cloud}}{\mathrm dr}=\frac{\dot{M}_{\rm grow}}{v_{\rm cloud}}-\frac{\dot{M}_{\rm loss}}{v_{\rm cloud}}.
    \label{cloud_mass}
\end{equation}
The cloud velocity equation consists of three terms including the momentum transfer between the wind and the cloud, the drag-term and a gravity term. Thus, the velocity evolution is given by
\begin{equation}
    \frac{\mathrm dv_{\rm cloud}}{\mathrm dr}=\frac{\dot{v}_{\rm cloud}}{v_{\rm cloud}}=\frac{\dot{p}_{\rm drag}+v_{\rm rel}\dot{M}_{\rm grow}-M_{\rm cloud}\frac{v_c^2}{r}}{M_{\rm cloud}v_{\rm cloud}}
    \label{cloud_velocity}
\end{equation}
with $v_{\rm c}$ the circular velocity related to gravity.
The metallicity equation that govern the system, are derived using terms only related to the exchange of metallicity between the phases:
\begin{equation}
    \frac{\mathrm dZ_{\rm cloud}}{\mathrm dr}=\frac{\dot{Z}_{\rm cloud}}{v_{\rm cloud}}=\frac{Z_{\rm wind}-Z_{\rm cloud}\dot{M}_{\rm grow}}{M_{\rm cloud}v_{\rm cloud}}
    \label{cloud_metallicity}
\end{equation}
Equations~\ref{cloud_mass}-\ref{cloud_metallicity} govern the cloud evolution in the dynamical wind background. In order to complete the system of differential equations for the whole cloud-wind model, we have to see how these equations affect the wind macroscopically.
\subsubsection{Wind related quantities}
The next step is to zoom out from the `microscopical' picture of the cloud-wind interaction, in order to connect the quantities derived in the last chapter with the source terms in the wind equations. We mainly focus on the final set of equations here; a more detailed derivation can be found in \citetalias{Fielding_2022}. 

The rate of how the wind gains or loses, in other words, the mass source term is related to how each individual cloud gives or drains mass from the wind, is given as
\begin{equation}
    \dot{\rho}=-n_{\rm cloud}\dot{M}_{\rm cloud}=-n_{\rm cloud}\dot{M}_{\rm grow}+n_{\rm cloud}\dot{M}_{\rm loss}=-\dot{\rho}_{-}+\dot{\rho}_+
\end{equation}
the momentum source term can be derived with the exact same procedure, adding a drag term due to the relative motion of the cloud in the wind:
\begin{equation}
    \dot{p}=-v_{\rm wind}\dot{\rho}_{-}+v_{\rm cloud}\dot{\rho}_{+}-\dot{p}_{\rm drag}
\end{equation}
with the drag-term being:
\begin{equation}
    \dot{p}_{\rm drag}=\frac{n_{\rm cloud}}{2}\mathcal{D}_{\rm drag}\rho_{\rm wind}v_{\rm rel}^24\pi r_{\rm cloud}^2
\end{equation}
and $\mathcal{D}_{\rm drag}$ is the drag coefficient.
The energy term is similar. Defining:
\begin{equation}
    \mathcal{V}\equiv\frac{v^2}{2}+\frac{\gamma}{\gamma-1}\frac{P}{\rho}-\frac{1}{2}v_{\rm esc}^2
\end{equation}
we can write:
\begin{equation}
    \dot{q}=-\mathcal{V}^2\dot{\rho}_{-}+\mathcal{V}_{\rm cloud}^2\dot{\rho}_{+}-\dot{p}_{\rm drag}v_{\rm cloud}
\end{equation}
This is the form of the source terms in the wind equations, caused by the interaction with a single cloud population. We generalise these source terms, by adding a number of different cloud populations with different initial masses. We ignore cloud-cloud interactions in our model, so no cross-terms will appear in the equations bellow. Thus, we have for the total mass, momentum and energy:
\begin{equation}
    \dot{\rho}=\sum_i\dot{\rho_i}=\sum_i n_{\rm cloud,i}\dot{M}_{\rm cloud,i}
    \label{source_mass}
\end{equation}
\begin{equation}
    \dot{p}=\sum_i \dot{p_i}=\sum_i n_{\rm cloud,i}\dot{p}_{\rm cloud,i}
    \label{source_momentum}
\end{equation}
\begin{equation}
    \dot{q}=\sum_i q_i=\sum_i n_{\rm cloud,i}\dot{q}_{\rm cloud,i}
    \label{source_energy}
\end{equation}
with the terms $\dot{p}_{\rm cloud},\dot{q}_{\rm cloud}$ constructed in detail in \citetalias{Fielding_2022}.

Each cloud population "i" has a different initial mass, and its number density is affected by the choice of distribution.
We use discretised versions of continuous probability distributions in the following way:
\begin{equation}
    \rm Prob(M_{\rm 0,cloud,\mathrm{i}})=\frac{d(M_{0,\rm cloud,i})}{\sum_\mathrm{i}d(M_{0,\rm cloud,\mathrm{i}})}
\end{equation}
with $d(M_{0,\rm cloud,i})$ the probability density function for a given distribution.  We assume that the initial cold mass fraction is picked up from a probability distribution:
\begin{equation}
    \eta_{\rm cold,i}=\eta_{\rm cold,t}\rm Prob(M_{0,\rm cloud,i})=\eta_{cold,t}\frac{d(M_{0,\rm cloud,i})}{\sum_id(M_{0,\rm cloud,i})}
\end{equation}
with $\eta_{\rm cold,t}$ the total cold mass fraction and $M_{0,\rm cloud,i}$ the initial mass for a cloud of the population "$i$".\\
By defining the initial cloud number flux to be:
 \begin{equation}
     \dot{N}_\mathrm{i,0}=\frac{\text{SFR}\cdot\eta_{cold,\mathrm{i}}}{\langle M_{0,\rm cloud}\rangle}=\frac{\text{SFR}\cdot d(M_{0,\rm cloud,\mathrm{i}})}{\sum_\mathrm{i}M_{0,\rm cloud,\mathrm{i}} d(M_{0,\rm cloud,\mathrm{i}})}
 \end{equation}
 The clouds are injected gradually between $R_{\rm disc}$ and $1.33 R_{\rm disk}$ with a powerlaw, which means:
 \begin{equation}
     \dot{N}_{\mathrm{i}}=\dot{N}_{\mathrm{i,0}}\left(\frac{r}{1.33R_{\rm disc}}\right)^\beta
 \end{equation}
 with $R_{\rm disc}=300\;\rm pc$ and $\beta=10$. The choice of $\beta$ can be tuned without significant changes.
Thus, we have for the number density of the clouds:
\begin{equation}
    n_\mathrm{i}=\frac{N_\mathrm{i}}{V_\mathrm{i}}=\frac{\dot{N}_\mathrm{i}}{\Omega r^2 v_{\rm cloud,\mathrm{i}}}
\end{equation}

Now all the quantities in the equations are specified, so we can go on solving the model for different conditions. 

\subsubsection{Multiple cloud populations}
In the general case of our model, the wind interacts simultaneously with different cloud populations, meaning groups of clouds that have initially different masses. The number of different cloud masses is arbitrary and is a free parameter in our model. The cold mass fraction for every cloud population comes along with a certain probability, and this probability is completely specified by the choice of the probability distribution the cloud masses follow.

We investigate three different examples:
\begin{itemize}
\item a lognormal distribution:
\begin{equation}
    d(M_{\rm0,cloud,i})=\frac{1}{M_{\rm 0,cloud,i}\sigma\sqrt{2\pi}}\exp{\left(-\frac{\left(\log M_{\rm 0,cloud,i}-\mu\right)^2}{2\sigma^2}\right)}
\end{equation}
with $\mu$ the mean of the distribution and $\sigma$ being free parameters of our model.

\item a powerlaw distribution:
\begin{equation}
    d(M_{\rm 0,cloud,i})\propto\frac{1}{M_{\rm 0,cloud,i}^{\rm b}}
\end{equation}
with $b=2$ and the range of the powerlaw distribution is a free parameter for the model (it is the range of the initial cloud masses we choose). Note that the proportionality factors do not need to be determined as the quantity is normalized by the sum of each value.

\item a delta-distribution:
\begin{equation}
    d\left(M_{\rm 0,cloud,i}\right)=\delta\left(M_{\rm 0,cloud,i}-M_{0}\right)
\end{equation}
with the singular cloud mass $M_0$ being the only free parameter. This distribution is identical to the single cloud mass case and allows us to study the impact of a more realistic cloud mass distribution.
\end{itemize}

\section{Results}
\subsection{Homogeneous background}
\label{subsection:homogeneous} 
First, we will study the case with a constant background medium. While this is not realistic for a galactic wind, it will allow us to compare our analytic model with hydrodynamical `cloud crushing' simulations \citep{Gronke_2018,2020MNRAS.492.1841L,Abruzzo_2022}
 
The differential equations that govern this system are the evolving rates of the cloud mass $\Dot{M}_{\rm cloud}$, velocity $\Dot{v}_{ \rm cloud}$, metallicity $\Dot{Z}_{\rm cloud}$ and the length of the cloud tail $\dot{L}$. So the equations involved are:
\begin{equation}
    \dot{M}_{\rm cloud}=2\pi r_{\rm cloud}\rho v_{\rm in}L- \frac{M_{\rm cloud}}{r_{\rm cloud}\chi^{1/2}\sqrt{1+\mathcal{M}_{\rm rel}}}\frac{v_{\rm rel}}{2}
\end{equation}
\begin{equation}
    \dot{v}_{\rm cloud}=\frac{\dot{p}_{\rm drag}+v_{\rm rel}\dot{M}_{\rm grow}-M_{\rm cloud}\frac{v_c^2}{r}}{M_{\rm cloud}}
\end{equation}
\begin{equation}
\dot{Z}_{\rm cloud}=\frac{Z_{\rm wind}-Z_{\rm cloud}\dot{M}_{\rm grow}}{M_{\rm cloud}}
\end{equation}
\begin{equation}  
\dot{L}=v_{\rm rel}
\end{equation}

This limited case even though not so physically interesting, provides a test for how the model can reproduce the results from small scale hydrodynamical (`cloud crushing') simulations. In this scenario, the equations are influenced by how the tail evolves, especially in the early stages of the evolution. More specifically, without including a cut-off to the tail length  (like Eq.~\eqref{minimum_length}), the growth of the tail -- even though the clouds initially lose mass -- leads to a regrowth of the cloud. This is true even for clouds losing almost all their mass ($M_{\rm cloud}< 0.05 M_{\rm cloud,init}$) after several $t_{\rm cc}$. This behavior is moderated by imposing the minimum on the length evolution which accounts to the fact that the length cannot be growing when the cloud is losing mass.

It is worth mentioning that the use of the cut-off does not rule out the possibility of the cloud to start growing again even after losing a significant part of its mass. This is an effect that is commonly observed in simulations and is possible in our model especially in the multicloud case. The role of the cut-off has the physical meaning of regulating the evolution of the tail, when the mass is not enough to allow its formation, and is also essential in order to reproduce the results of simulations such as \citet{Gronke_2018}.  Furthermore, in order to have the best agreement with the runs of the above hydrodynamical simulations we impose a fudge factor $a=2$ for the $\dot{M}_{\rm loss}$ term in Eq.~\eqref{mass_loss}.

\begin{figure*}
	\includegraphics[width=0.9\linewidth]{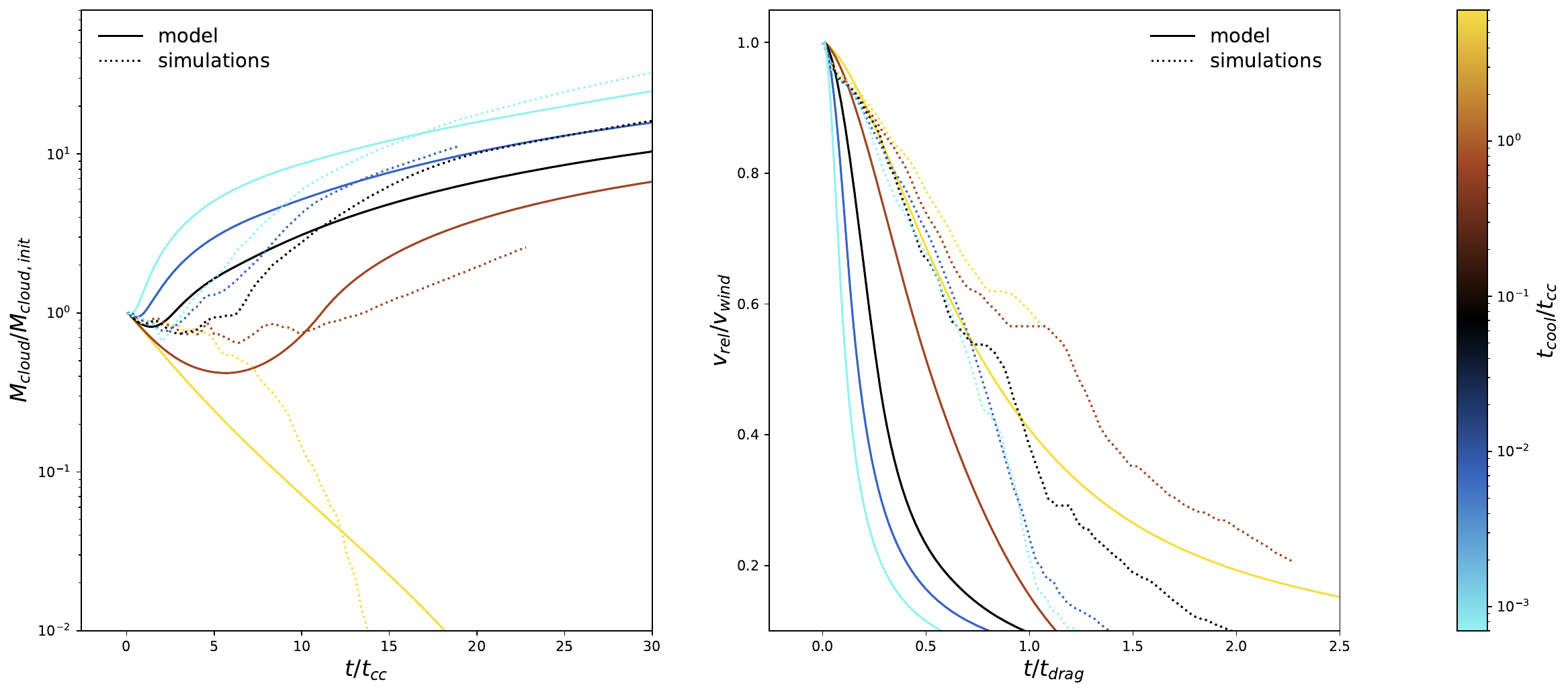}
    \caption{The time evolution of the cloud mass is presented in the left hand panel, while the time evolution of the relative velocity is shown in the right hand panel. The solutions from the model of the paper are presented in straight lines in comparison with the results of hydrodynamic simulations from \citet{Gronke_2018} which are shown in dashed. In order to use the same conditions as the hydrodynamic simulations, we use $\chi=100, T_{\rm cloud}=4\cdot10^4\text{K}$ and  wind Mach number $\mathcal{M}=1.5$. The Pressure of the wind is $P_{\rm wind}/\text{kb}=10^4 \text{K}\;\text{cm}^{-3}$. Different initial values for $t_{\rm cool,mix}/t_{\rm cc}$ are presented with different colors.} 
    \label{fig:comparison_chi_100}
\end{figure*}

\begin{figure}
	\includegraphics[width=1.\linewidth]{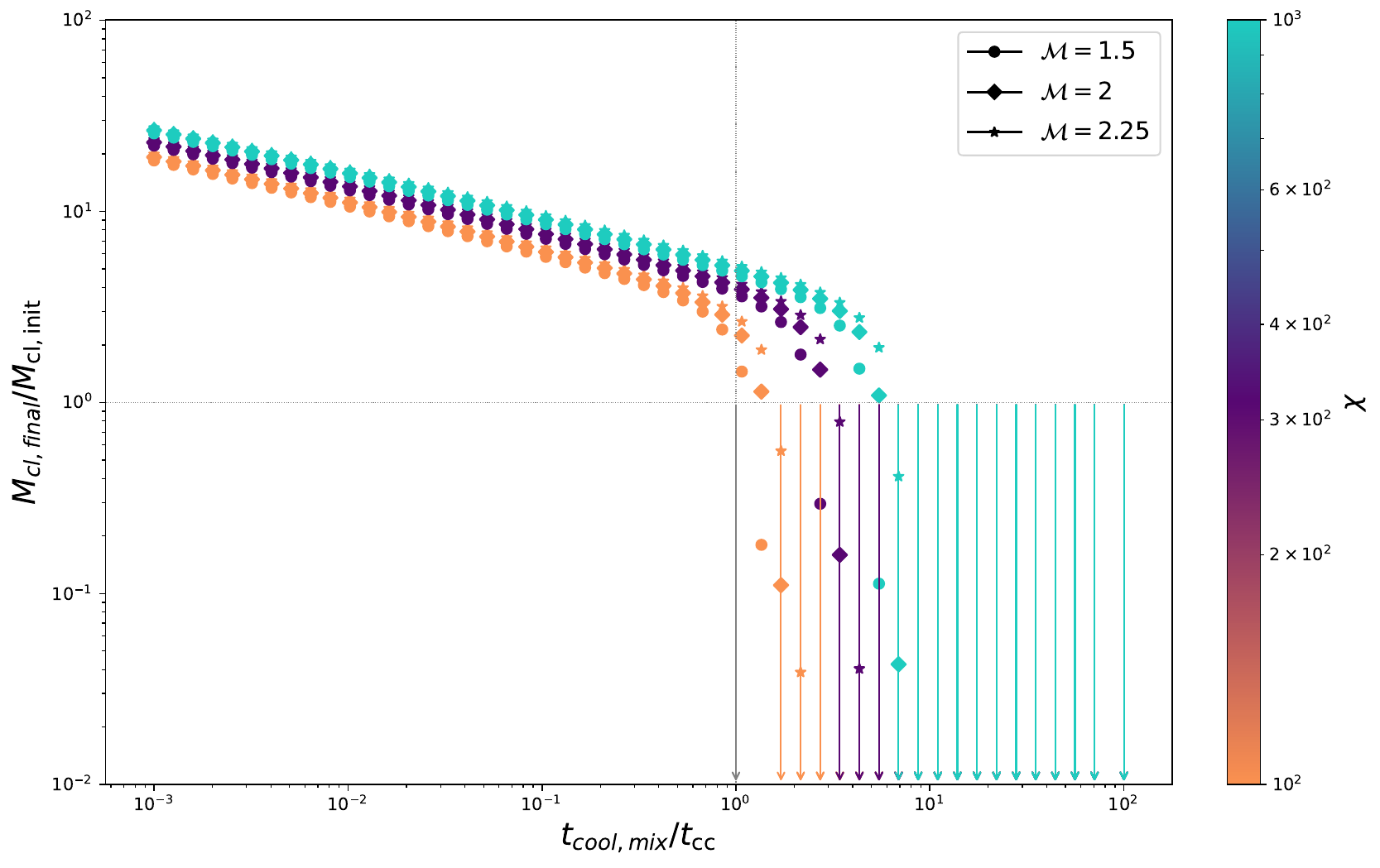}
    \caption{A scatter plot presenting the final values of the cloud masses, normalised to the initial ones, as a function of the $t_{\rm cool,mix}/t_{\rm cc}$ criterion. The different colors correspond to three different $\chi=100,300,1000$ and the symbols to 3 different wind Mach numbers $\mathcal{M}=1.5,2.25,2.5$. The allowed regions by the criterion are the upper left box and the lower right one. } 
    \label{fig:scatter-uniform}
\end{figure}

From Fig.~\ref{fig:comparison_chi_100}, we see that the solution of the equations match qualitatively well with the results of hydrodynamic simulations carried out by \citet{Gronke_2018}.We see that the evolution behaviour between the model and the simulations is matched (regarding wether the cloud survives or not for different sets of initial conditions). Even though we do not have a one to one correspondence with the simulations, this is generally not the purpose of models like this. In this figure, we show the case of $\chi=100$ and  we present  higher overdensities in Appendix~\ref{app:different_chis} (cf. Fig.~\ref{fig:comparison300} and Fig.~\ref{fig:comparison1000} for $\chi=300$ and $\chi=1000$, respectively).

Except for small disagreements (e.g., for $\chi = 1000$, $t_{\rm cool,mix}/t_{\rm cc}=5$), we find that our model reproduces the behaviour of the mass evolution (shown in the left panel) reasonably well. For the case of the velocity (right panel), we see that the values for the model are generally lower. This is expected, as we overestimate the mass evolution, and due to momentum exchange, this leads to an underestimation of the velocity.
We observe, for instance, that in our model, as well as in the simulations, clouds initially tend to lose mass and then start regrowing again, with a tendency to keep growing as the evolution continues. This behavior, as mentioned earlier changes continuously as we vary the parameter range, so it happens even in the cases that clouds lose almost all their mass rapidly. This effect is capped in the cloud-destruction case by the limit for the evolution of the tail, which forces the clouds that are instantly shredded to not be able to form a tail and eventually regrow.

Figure~\ref{fig:scatter-uniform} shows the result of the model for a larger parameter range. Specifically, we show the mass after $30t_{\rm cc}$ as a function of $t_{\rm cool,mix}/t_{\rm cc}$ for various overdensities and Mach numbers. 
One can see that our model agrees with the survival criterion of \citet{Gronke_2018} within an order of magnitude.

There's a disagreement for larger overdensities where the models predicts mass growth even for $t_{\rm cool,mix}/t_{\rm cc}>1$, however, for $\chi > 10^3$ only very few hydrodynamical simulations have been run thus far (see, e.g., \citealp{abruzzo2022taming}) and how the survival ratio potentially changes for such overdensities is yet unclear. We, thus, fix $a=2$ for the remaining parts of this study and generalize next to a dynamical wind background.

\subsection{Single cloud population}
We proceed to increase the complexity of the model by allowing cloud-wind interactions with only one cloud population. This is the immediate next step, before moving on to the full multi-cloud case. 

The simplicity of the single cloud approach will allow us to understand better the behavior and the properties of the solutions. This will help us separate between the effects of individual cloud evolution and collective effects from the introduction of distributions in the next sections. Furthermore, we can compare the results of our model with recent work \citepalias{Fielding_2022}, as we will discuss in more details in Section~\ref{subsection: comparison-fielding}.

While the single case is presented as a limiting case here, the results are actually included and used in the multicloud case of section~\ref{subsection:multiplecloudssection}, as they are equivalent with using a delta distribution. A systematic study of this distribution is motivated by the fact that the behavior of the individual bins from this section can guide our intuition to the more general solutions, and will be used as a comparison to the results of different distributions.

The model now includes the wind evolution, so the initial conditions for our equations are determined by the \citet{Chevalier1985} model inside the galactic disc, and eventually the loading factors $\eta_{\rm mass,hot},\eta_{\rm mass,cold},\eta_{\rm energy}$ as in Eq.~\ref{eta_M}, as well as the value of the Star Formation Rate. The use of these initial conditions help us connect the model to the galaxy characteristics, making it easier to compare with observables. The clouds are introduced gradually outside the galactic disc. We move on to present the solution for the equations of the model in two different cases.
\begin{figure*}
	\includegraphics[width=0.9\textwidth]{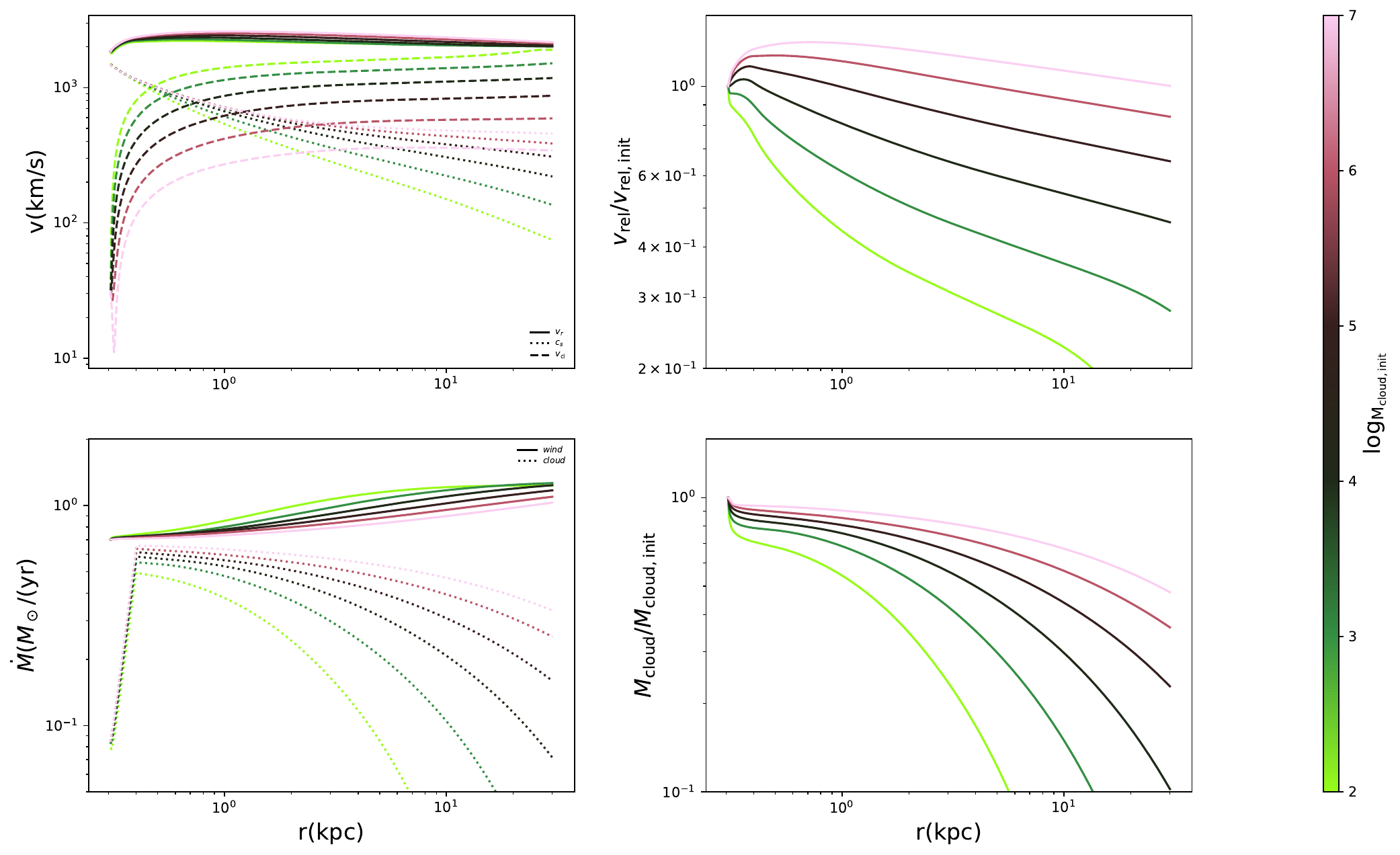}
    \caption{A four figure panel for the evolution of basic quantities for different initial cloud masses in the single cloud case. The colors correspond to the initial cloud masses, which range from $10^2(M_\odot)$ to $10^7(M_\odot)$. The initial conditions are $\text{SFR}=7\rm M_\odot/yr$, $\eta_{\rm mass,hot}=0.1$,$\eta_{\rm mass,cold}=0.1$,$\eta_{\rm energy}=1$.} 
    \label{fig:single-cloud1}
\end{figure*}
\begin{figure*}
	\includegraphics[width=.9\textwidth]
      {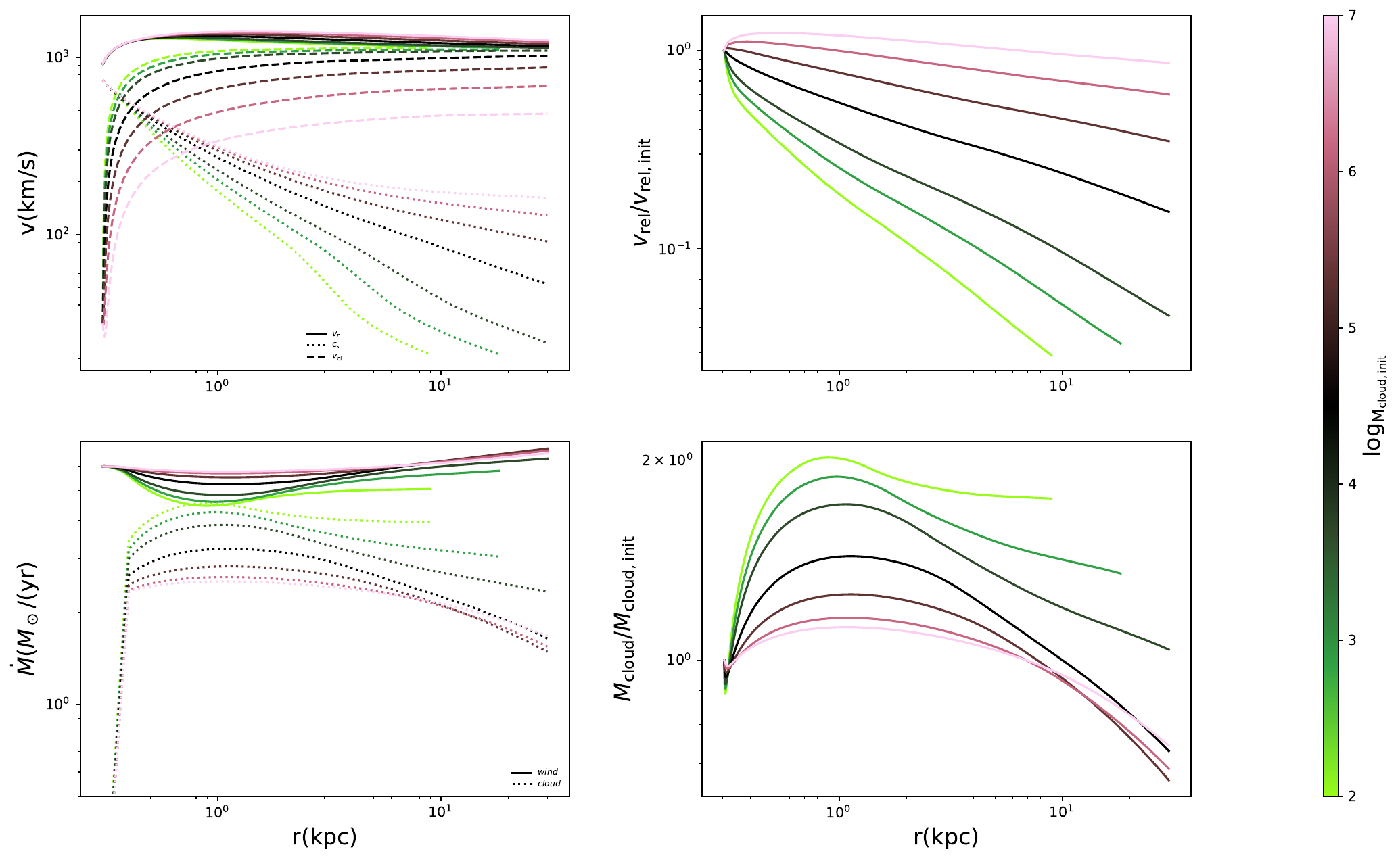}
    \caption{In this panel the initial masses range from $10^2(M_\odot)$ to $10^7(M_\odot)$. The initial conditions are $\text{SFR}=15\rm M_\odot/yr$, $\eta_{\rm mass,hot}=0.4$,$\eta_{\rm mass,cold}=0.15$,$\eta_{\rm energy}=1$. The integration ends sooner for the lightest cloud because the wind has cooled down to the cloud temperature.} 
    \label{fig:single-cloud2}
\end{figure*}

In Figure~\ref{fig:single-cloud1}, we display a scenario with lower star formation rate and loading factors. This is a case where clouds are destroyed, and there is mass loss in the whole integration interval. From the plots one can see that clouds are shredded instantly after they are introduced. As it is expected from the cloud evolution, the lightest clouds are shredded faster. As we move on to higher initial masses, heavier clouds lose mass more slowly and end up with around half of their mass. Even the heaviest clouds in this case are not able to grow or even retain their mass. 

 The next case is presented in Fig.~\ref{fig:single-cloud2}. Here we use larger SFR and $\eta_{\rm mass,hot}$. One would expect that the clouds would survive in this regime, because values for the initial conditions are higher (see \citetalias{Fielding_2022}). In reality we still see mass loss, even though the model exhibits a different behavior. Lighter clouds initially gain mass rapidly, and then start to lose mass again, because the $\dot{M}_{\rm loss} \propto v_{\rm rel}$ dominates over the $\dot{M}_{\rm grow}\propto v_{\rm in}$ term. This change is related again to the scalings of the inflow velocity. For the heavier clouds, we observe a destructive behaviour, but there are not significant differences between the initial and final value of their mass . For the lightest cloud $M_{\rm cloud}=10^2 M_{\odot}$, the integration stops because the wind has cooled down to the cloud temperature. This is a common feature for the high SFR cases, and only occurs for light clouds that accelerate rapidly. In general, we observe a significant difference in the behavior of the lightest clouds for the different initial conditions. Mainly there is the case exhibited before with rapid destruction and the one here, where clouds first grow and then lose mass. Heavier clouds do not follow this pattern, and their evolution is in general more steady. This leads us to check the model for the whole parameter range.
 \begin{figure}
	
	\includegraphics[width=0.9\linewidth]{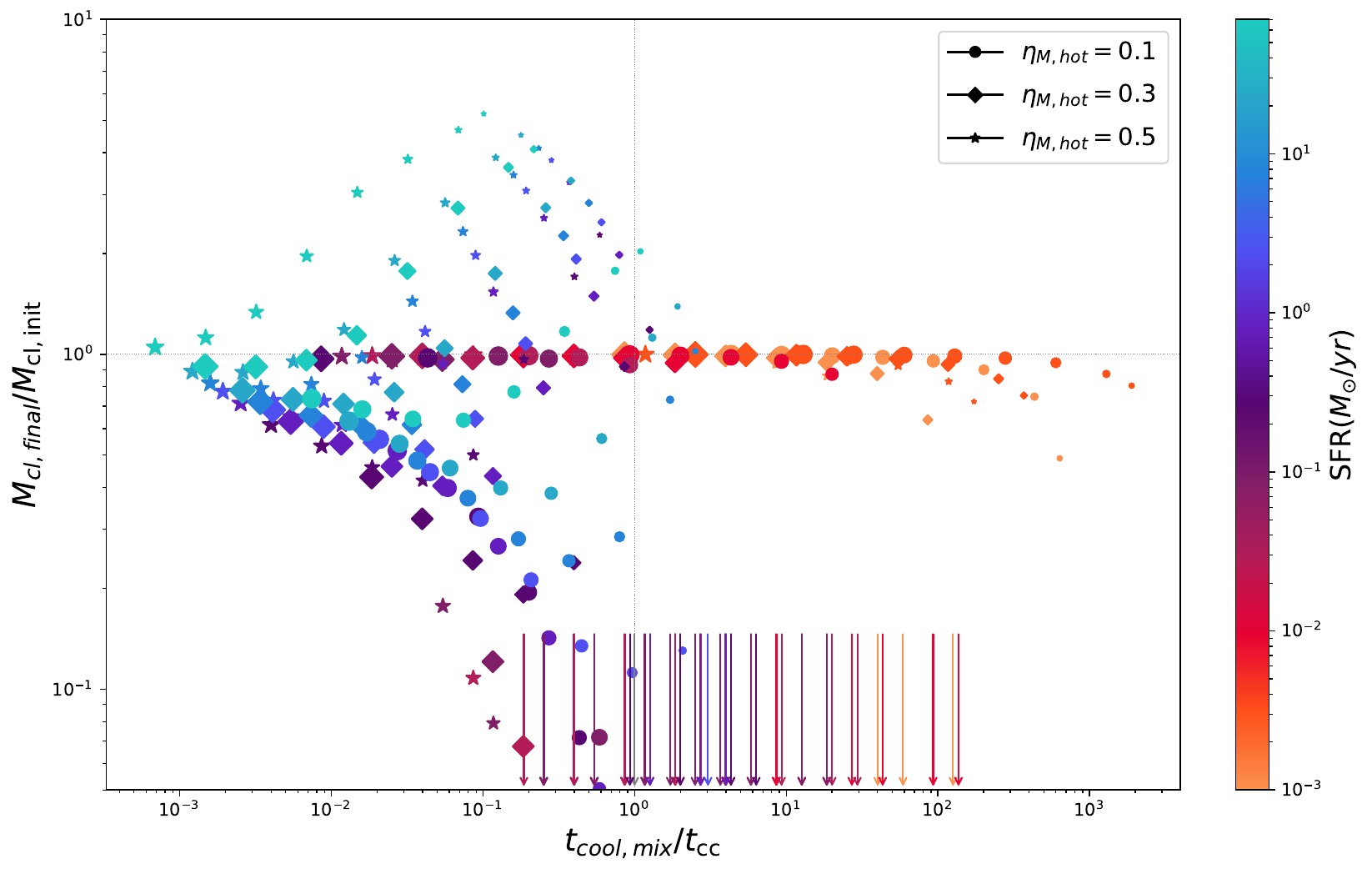}
    \caption{Overview of different wind solutions showing the final cold mass versus $t_{\rm cool,mix}/t_{\rm cc}$ evaluated at the cloud injection point $r=300pc$ (see Appendix~\ref{app:criterion} for different radii). The final cold mass is computed at $r\sim 30\rm kpc$. The initial cloud masses range from $10^0.5(M_\odot)$ to $10^7(M_\odot)$ and every marker's size is proportional to the corresponding mass. The Star Formation Rate ranges from SFR=$0.001(M_\odot/\rm yr)$ to $70(M_\odot/\rm yr)$ and the $\eta_{\rm mass,hot}=[0.1,0.3,0.5]$ } 
    \label{fig:scatter-single}
\end{figure}
 In Figure~\ref{fig:scatter-single}, we check the final value for the mass of the clouds at a radius of $r=30\rm kpc$ for different values of the survival criterion of \citet{Gronke_2018}, and for a bigger parameter range. We see that we have a general cloud destruction behaviour. As explained heavier clouds usually end up with a mass near their initial mass, as their mass evolution is really slow. Light clouds are more sensitive to the choice of the initial conditions. The case that light clouds end the integration interval with mass higher than their initial value does not mean survival. As we have seen in individual cases, the loss term dominates at some point in the interval, or the wind fails to sustain them. From this plot, we also see that we are in general out of the allowed regions for the survival criterion. We can conclude, that at least for our model, there is no indication to assume that the survival criterion of a homogeneous background holds for a dynamical wind.

A conclusion from the above paragraphs is that cloud-wind interactions are present, we see that the behavior of the model changes. Now, the clouds have a tendency to eventually die in the integration interval for almost all the parameter range of the initial conditions.   This is related to the saturation of the inflow velocity, which scales as $\sim c_{\rm s,h}$ as we can see from Eq~\ref{vturb-saturation}, while the loss term is always proportional to $\sim v_{\rm rel}$.

\subsection{Comparison with Fielding \& Bryan (2022)}
\label{subsection: comparison-fielding}
\begin{figure*}
	\includegraphics[width=0.9\textwidth]{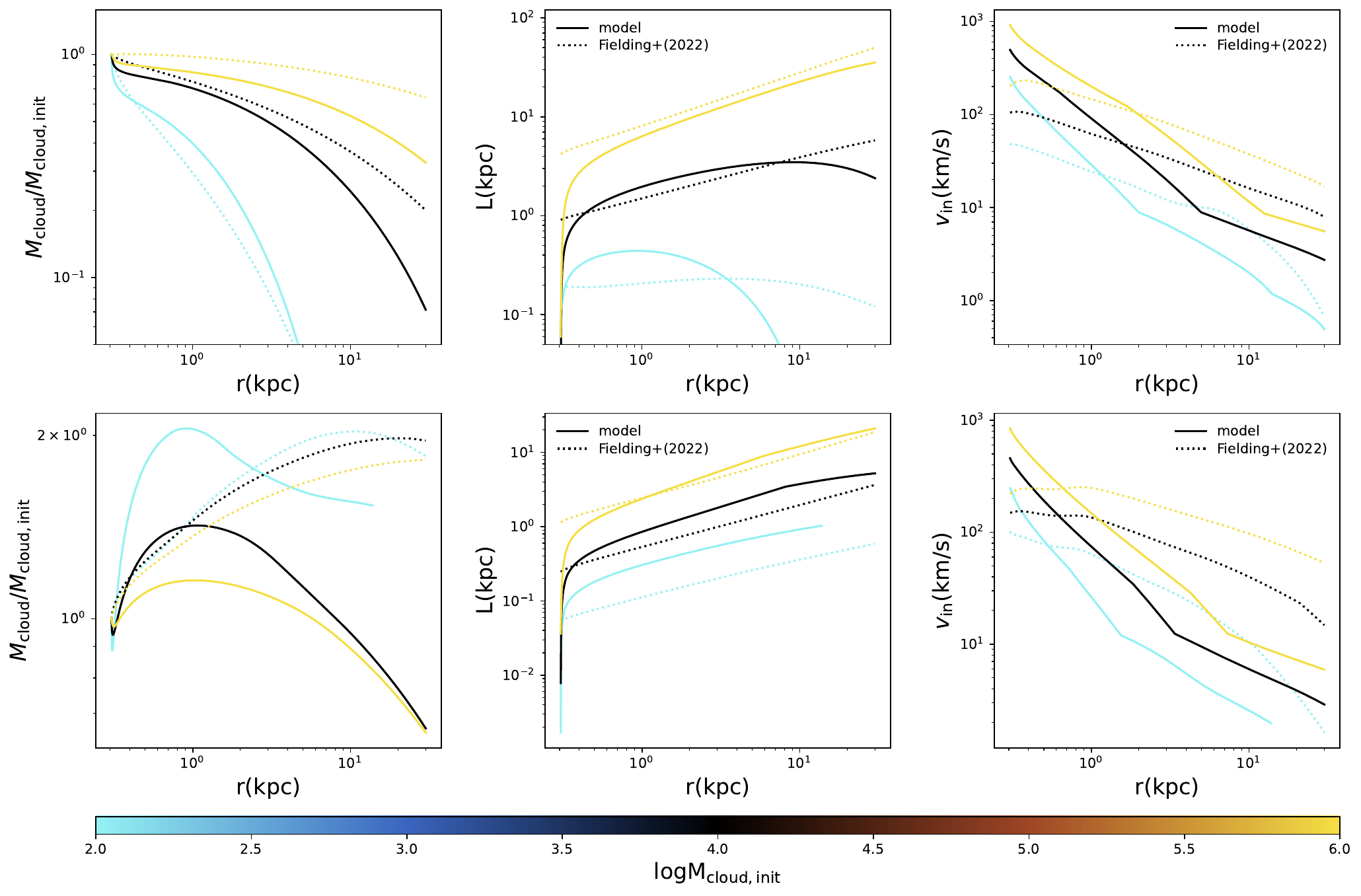}
    \caption{In this panel, we present the cloud mass, the length of the tail and the inflow velocity $v_{\rm in}$ as a function of the radius. The straight lines correspond to our model and the dashed ones correspond to \citet{Fielding_2022}, while the different colors correspond to different initial cloud masses. 2 different cases are presented here, with the first row corresponding to $\rm SFR=5 M_{\odot}/yr$, $\eta_{\rm mass,hot}=0.1$, $\eta_{\rm mass,cold}=0.1$, and the second row to $\rm SFR=10 M_{\odot}/yr$, $\eta_{\rm mass,hot}=0.5$, $\eta_{\rm mass,cold}=0.1$. The integration ends sooner for the cloud with $M_{\rm cloud,init}=10^2 M_{\odot}$ in the second row, because the wind has cooled down to the cloud temperature.} 
    \label{fig:comparison1}
\end{figure*}
In this single cloud case, we can compare our findings with similar models such as the one proposed by \citetalias{Fielding_2022}. The models differ in the choice of the cloud source terms as well as the length evolution equation and the equation for the $\dot{M}_{\rm cloud,loss}$ term. 

The changes in the cloud source terms eventually lead to a different inflow velocity $v_{\rm in}$ which is the term affecting the mass growth equation. The form we propose in our model, and specifically the dependence of the velocity to different physical quantities (see Eq~\ref{inflow_weak},\ref{inflow_strong}), is similar to the one from \citepalias{Fielding_2022} . The difference between the two is by using $t_{\rm cool,cold}$ instead of $t_{\rm cool,mix}$ and by using the exact scalings from the hydro-simulations of \citet{Tan_2021}. The use of the same scalings leads to small differences in the prefactors. Both of these changes contribute slightly to the behavior of the inflow velocity. The main difference in the behavior of the function comes from the cut off and the saturation we introduced in Equation~\ref{vin-min} and Equation~\ref{vturb-saturation}. This different behavior is what determines whether clouds get shattered or not between the two models.

The tail growth function has a more natural evolution, governed by a dynamical Equation~\ref{expansion-equation-radial}. The tail in the compared model is assumed to be already formed when the clouds are introduced. This assumption gives higher values for the cooling area from the beginning, boosting the cloud growth term. 

The terms related to loss for the cloud mass have similar functional form, but differ by pre-factors. The fiducial prefactors of \citetalias{Fielding_2022} might result in difference in at most one order of magnitude between the models, depending on the choice for the value of the prefactor. This effect individually contributes slightly to the general results, but is one of the factors for the difference between cloud survival or not between the models. Furthermore, our equation for loss includes the delayed destruction timescale of \citet{Scannapieco_2015}.

Another difference that contributes to these results is that in the second model, mass evolution is determined by a $\xi$ factor that differs by the $\frac{t_{\rm cool,mix}}{t_{\rm cc}}$ criterion for $\chi>100$. This will lead to differences as the common initial conditions for the  dynamical wind background usually gives values $\chi>1000$. The mass evolution equations are built to depend directly on this $\xi$ factor, which will lead to different survival criteria between the two models. In our case, there is no `built in' factor that determines the evolution, with the $t_{\rm cool,mix}/t_{\rm cc}$ being satisfied rather naturally in the homogeneous case.

In Figure~\ref{fig:comparison1} we present the differences between the models for the cloud mass, inflow velocity and tail evolution for two different scenarios. The upper series of plots correspond to the instant cloud dying scenario, while the lower to a case with higher SFR. In the second case the two behaviours differ. We see that in a dying cloud scenario, the behaviour of the masses is similar. However, for a case with higher SFR, $\eta_{\rm M,hot}$, the second model achieves cloud survival. The length of the tail in our case evolves and asymptotically approaches similar values as the other model for higher radii. The inflow velocities in our model always start with higher values but quickly their values become lower. In our case, we can see the two different cappings for the inflow velocities as introduced in Equations~\ref{vturb-saturation},\ref{vin-min}.

It can now be seen clearly that the main differences between the models that lead to the different results are the different inflow velocities used. If we focus on the lower panels in Fig.~\ref{fig:comparison1}, it is interesting to observe that mass loss in our model  always starts near the point where the inflow velocity gets lower than the one from \citetalias{Fielding_2022}. Because the most important difference between the two inflow velocities is the bounds used, the destructive behaviour in our case is connected to these terms.

\subsection{Multiple cloud populations}
\label{subsection:multiplecloudssection}
After exploring the various limiting cases, we can now check how the model behaves in its full form, using multiple cloud populations, with the initial mass of each population determined by different probability distributions.

As mentioned in earlier sections, we use three different probability distributions, a log-normal, a power-law ($d\propto 1/M_{\rm cloud,init}^2$ with a varying cut-off ranging from $[M_{\rm min},10^7 M_{\odot}]$. The values for the lower bound change from $M_{\rm min}\in[10^2M_{\odot},10^7M_{\odot}]$ depending on the mean mass of the distribution we choose.) and a delta distribution (which is the same as the single cloud case). The model can be run with an arbitrary number of discrete cloud populations $N$ sampled from the respective distributions. We confirmed that increasing this number from our fiducial value of $N\sim 20$ does not change the results significantly. Because in our model clouds do not interact with each other, the effect of the distribution affects the wind, as there is a sum of number densities for the clouds that is included in the equations, producing a collective effect on the wind quantities. This effect then back reacts on the clouds affecting their evolution. The effect of the distributions to the behaviour of the model depends on the initial conditions we choose. 

Examining the total mass for all populations as a function of radius, leads to ambiguities as they move with different velocities in the wind medium. We check the total ratio of the cold mass flux in the end of the integration interval for different values of the mean mass initially. By doing this, we investigate the tendency of whether the total cold mass available in the system will grow or not, when we change the mean mass.
\begin{figure*}
	\includegraphics[width=0.9\linewidth]{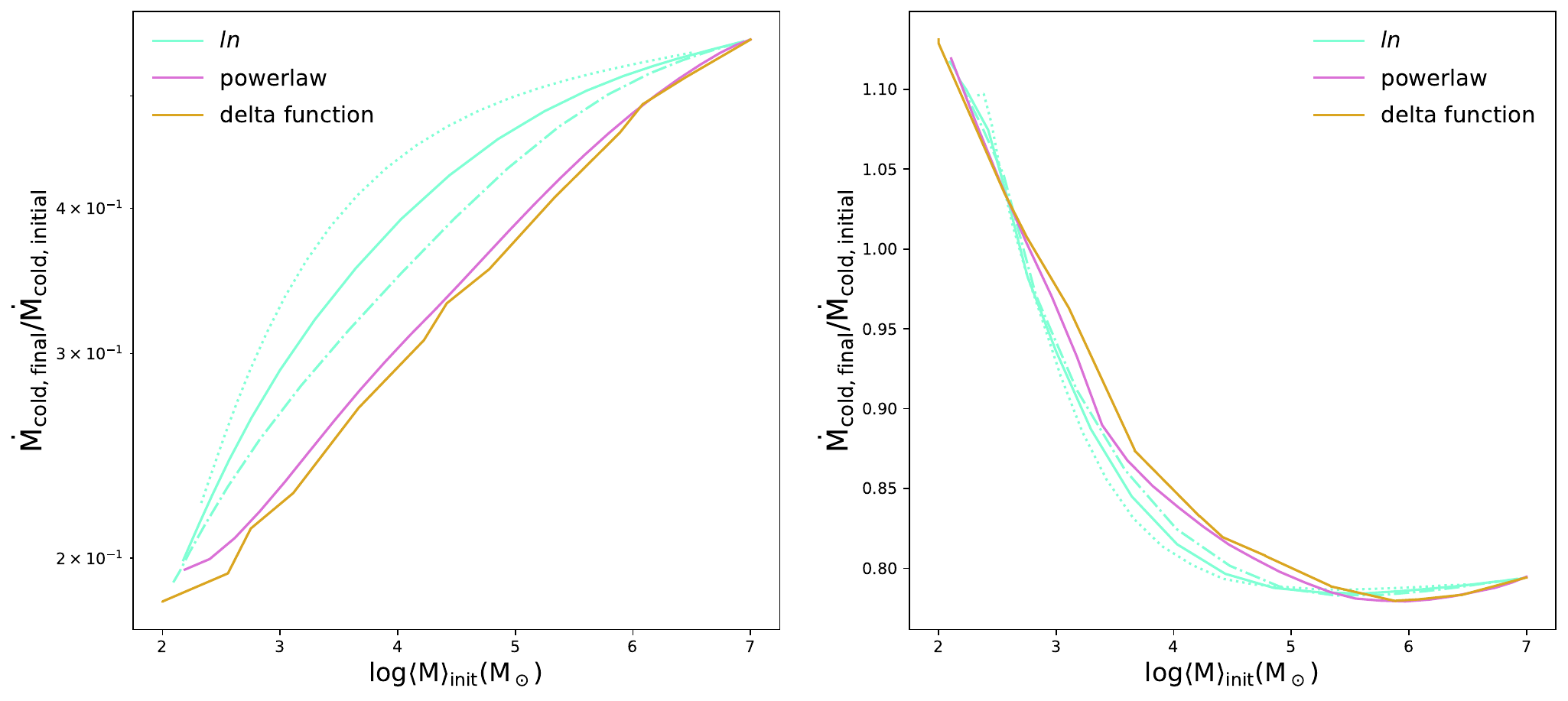}
    \caption{We present the ratio of final cold mass flux over the initial value after the clouds are introduced as a function of the mean initial mass of the clouds populations. The different colours correspond to lognormal, powerlaw and delta distributions. The straight $\sigma=1$ and the dashed lines correspond to $\sigma=0.75,1.5$ for the lognormal distribution. The left plot corresponds to the case where $\rm SFR=5 M_{\odot}/yr$, $\eta_{\rm mass, hot}=0.1$, $\eta_{\rm mass, cold}=0.1$, while the right has : $\rm SFR=25 M_{\odot}/yr$, $\eta_{\rm mass, hot}=0.5$, $\eta_{\rm mass, cold}=0.15$. }
    \label{fig:multicloud_mean_mass1}
\end{figure*}
In Figure~\ref{fig:multicloud_mean_mass1}, we explore two different parameter regimes, with the left plot corresponding to a case, where all clouds die. We observe that by increasing the mean initial mass of the cloud populations, the total cold mass flux tends to increase. Here a log-normal distribution seems to help the total cold mass lose mass more slowly, while the power-law and delta do not have significant differences. The similar behavior between these two is expected, as the leading contribution for the power law is from masses only near the mean value, exactly as in the $\delta$-distribution. As we have also seen in the single cloud case, here different initial masses, have significantly different evolution regarding the way they lose mass. Even one order of magnitude difference in the initial mass leads to an accountable difference in the percentage of mass they lose finally. When we choose a lognormal distribution, we allow these differences to contribute to the final outcome, leading to the difference from the other two distributions. By making the lognormal wider ($\sigma=1.5$), the effect is even more enhanced, as expected as we allow more and more populations to contribute, and the mass loss becomes slower.

In the right panel of Fig.~\ref{fig:multicloud_mean_mass1}, we show an intermediate regime where some clouds survive while others do not. We observe that for heavier mean mass initially, the flux ratio decreases. This can be explained by the fact that small clouds gain mass faster than the heavy ones. However, we see that we have again a slight increase  for higher mean mass values. In this case almost all clouds gain more mass initially in the interval and then start to lose mass again. The general behaviour is that while the heaviest clouds gain mass with the lowest rate, they also lose it with the lowest rate. So in our case, when all clouds enter the losing phase, the heaviest clouds start to lose mass with the lowest rate, and end up the interval with more mass (normalised to their initial value) than the next lighter ones. The differences between the distributions here are not significant. This behavior is expected, by analysing the evolution of each population. Here the total differences between the final and initial mass are not big, with clouds ending up the integration with slightly more or slightly less of their initial mass. Because of this, even if we use distributions that allow a wide range of masses to contribute, the total effect will not be significantly different, because we do not have big variations in the evolution between the populations.

\begin{figure}
	\includegraphics[width=0.9\linewidth]{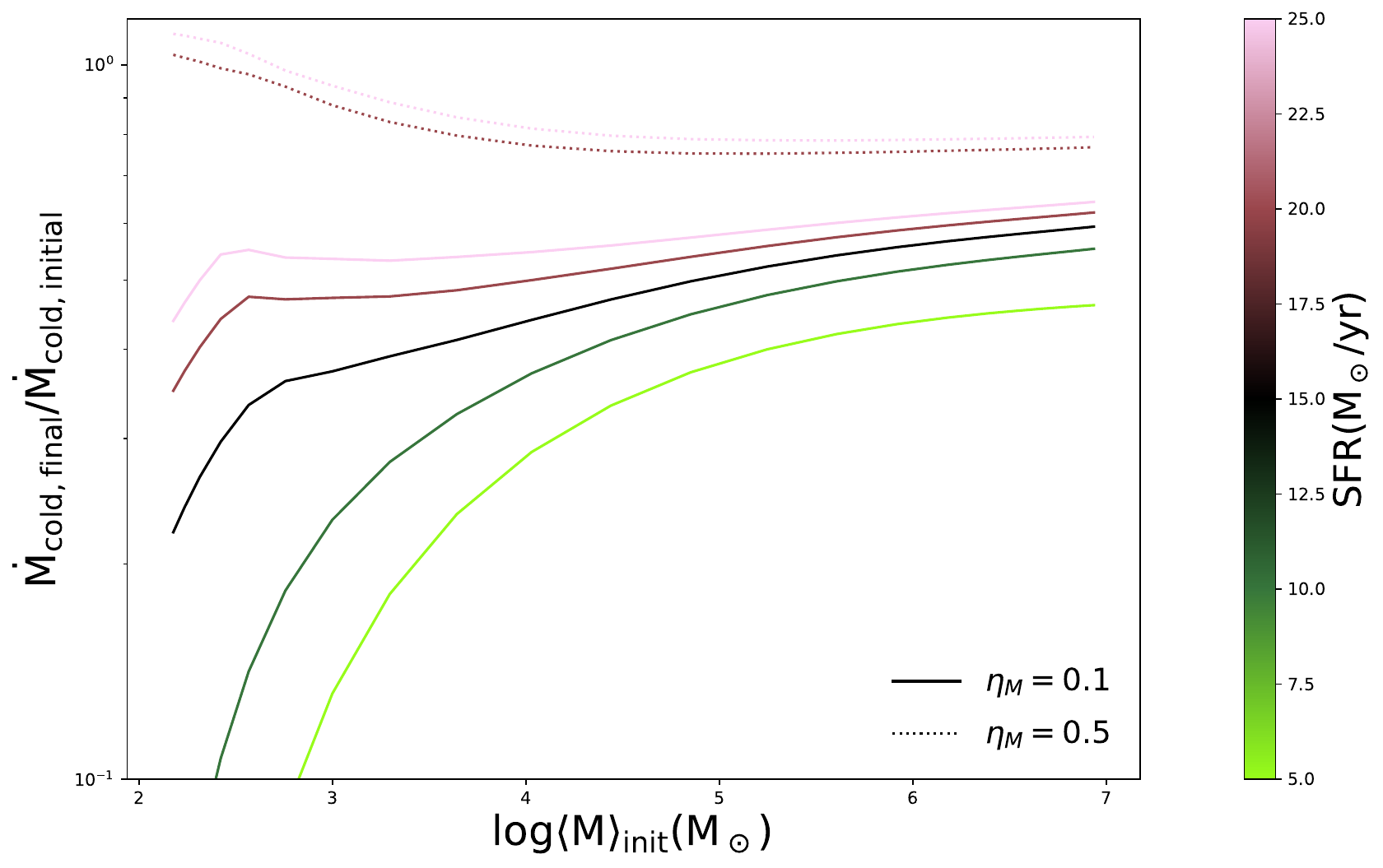}
    \caption{In this case, a lognormal distribution with $\sigma=1$ is picked. The behaviour of the flux ratio is again presented with the different colors corresponding to different star formation rates, while dashed lines correspond to different $\eta_{\rm mass, hot}$, while $\eta_{\rm mass,cold}=0.15$ remains constant. The highest curve in this plot corresponds to the highest lognormal curve in Fig.~\ref{fig:multicloud_mean_mass1}.
    }
    \label{fig:multicloud_mean_mass2}
\end{figure}
In Fig.~\ref{fig:multicloud_mean_mass2}, we pick a log-normal distribution with $\sigma=1$ and explore how the behavior changes for different initial conditions. The lowest curve corresponds to the log normal curve in the left panel of Fig.~\ref{fig:multicloud_mean_mass1}, while the highest to the one in the right panel. Here we can see exactly how with increasing the SFR and the $\eta_{\rm mass,hot}$ we move from a case where all clouds are shredded from the beginning, to the case where at least the lighter clouds tend to grow initially and end the integration interval with higher masses even though they have the tendency to die. From this plot we also see the trends for the evolution of different masses. Lighter clouds have significant differences for different initial conditions from instantly shredded to initial growth, while the heavy ones have a very steady behavior which either tends to be the same or does not differ by orders of magnitude. This sensitivity of the lighter clouds to initial conditions is a clear demonstration of the argument that heavy clouds have a steady evolution in contrast to the lighter ones.

The results presented in these plots are not specific to the chosen initial conditions. Mainly, for a case with instant shredding, there is a big difference on the amount of mass each population retains. When moving to cases with initial growth, there are not big variations on the amount of mass the clouds gain or lose.

Starting from the second scenario, the collective effect produced is as having a single cloud with the same mean mass, so the choice of the distribution does not really affect the results. The reason for this behavior is that because each cloud mass does not have significant changes in its evolution, so by including more cloud masses in the interaction does not lead to a large effect on the wind and then back to the clouds.

On the other hand, we observe cases where the choice of distribution plays a significant role to the results of the model. This is the first scenario, where the evolution of each cloud population has a big difference between different initial cloud masses. So when we use distributions, we include contributions from different masses, thus resulting in an observed effect to the results. 

In general, if we focus on the individual behaviour of one cloud of each population we again observe the heavier clouds have a steady evolution, while the lighter clouds have bigger changes in their evolution along the integration interval. What is different with the single cloud case is that sometimes, if a large range of initial cloud masses is included in the model, the rapid growth of small clouds is enhanced. Specifically, we observe cases where lighter clouds tend to regrow with the presence of heavier cloud populations. The contribution of this effect to the whole system depends on how far are the lightest populations from the mean value. Because of this, it is difficult to see the effects when total quantities are examined. Apart from this scenario, the lighter clouds either are shredded instantly, or gain mass rapidly. The loss term always dominates leading to destruction, even if they end up with higher mass than their initial at the end of the interval. So even if the behavior of each individual cloud might not be exactly the same as the single cloud case, we do not expect large differences when we examine mean values for the model, regarding the general behaviour.

\subsection{Lyman-alpha Halo Surface Brightness}
\label{subsection:observations}
Due to the cooling processes taking place at the cloud-wind interface, we have $\text{Ly}\alpha$ emissions from our model. We calculate  surface brightness along the line of sight as well as the total luminosity emitted, to connect our model with observations. One can relate the mass transfer rate from the hot to the cold medium (of a given cloud population $i$) to a total luminosity via \citep[e.g.,][]{Ji_Oh_Masterson_2019,Gronke_Oh_Ji_Norman_2021}
\begin{equation}
    \Tilde{L}_i=\frac{5}{2}\frac{k_{\rm b}T_{\rm hot}\left(\mathcal{M}^2+1\right)\dot{M}_{\rm grow}}{\mu m_{\rm p}}
\end{equation}
where $k_b$ is the Boltzmann constant and $m_{\rm p}$ is the mass of the proton
with 
\begin{equation}
    \mathcal{M}=\frac{v_{\rm turb}}{c_{\rm hot}}.
\end{equation}
The total luminosity is therefore given by
\begin{equation}
    L_{\rm total}=\sum_i \int dr 4\pi r^2 n_{\rm cloud,i}\Tilde{L}_i
\end{equation} 
with 
\begin{equation}
    \epsilon(r)=\sum_i n_{\rm cloud,i}\Tilde{L}_i
\end{equation}
being the emissivity. This implies that the surface brightness at an impact parameter $b$ from the galactic center is
\begin{equation}
    SB(b)=2\int_b^R dr \frac{\epsilon(r) r}{\sqrt{r^2-b^2}} = 2\int_b^R dr \frac{r n_{\rm cloud,i}\Tilde{L}_i }{\sqrt{r^2-b^2}}.
\end{equation}

\begin{figure}
	\includegraphics[width=0.9\linewidth]{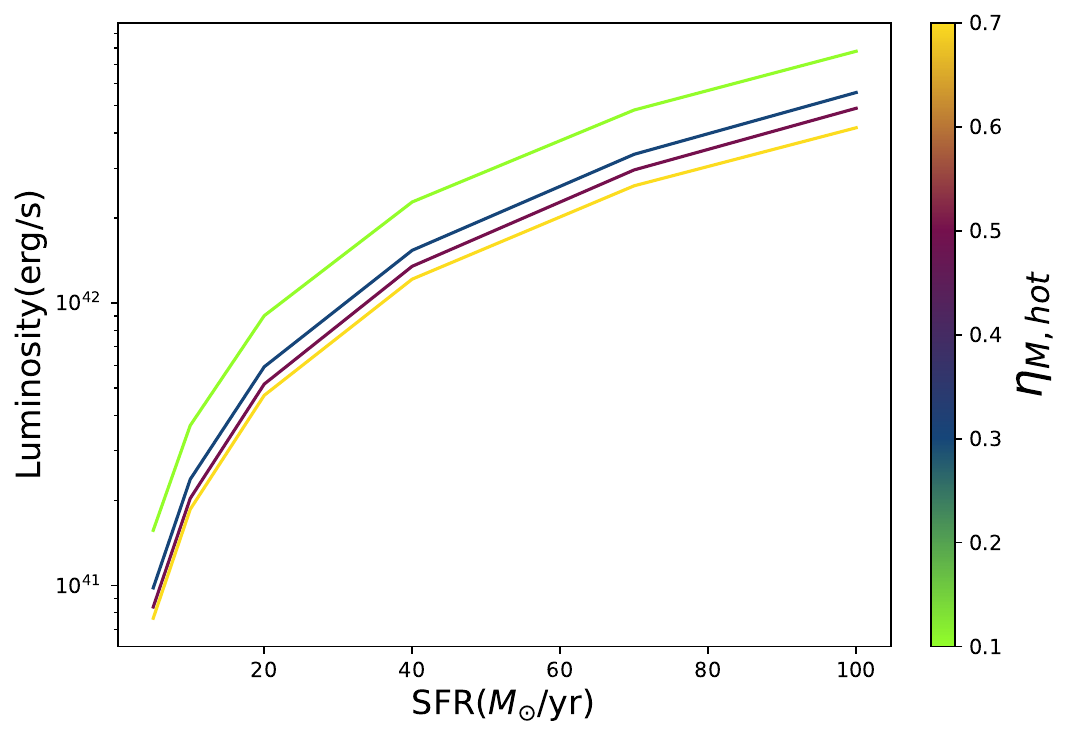}
    \caption{The halo luminosity as a function of the Star formation rate, with different colors corresponding to different $\eta_{\rm mass, hot}$. The case studied here is a lognormal distribution with $\sigma=1$ and mean initial mass $\rm \sim 10^{5.5} M_{\odot}/yr$} 
    \label{Luminosity1}
\end{figure}

\begin{figure*}
\includegraphics[width=0.9\linewidth]{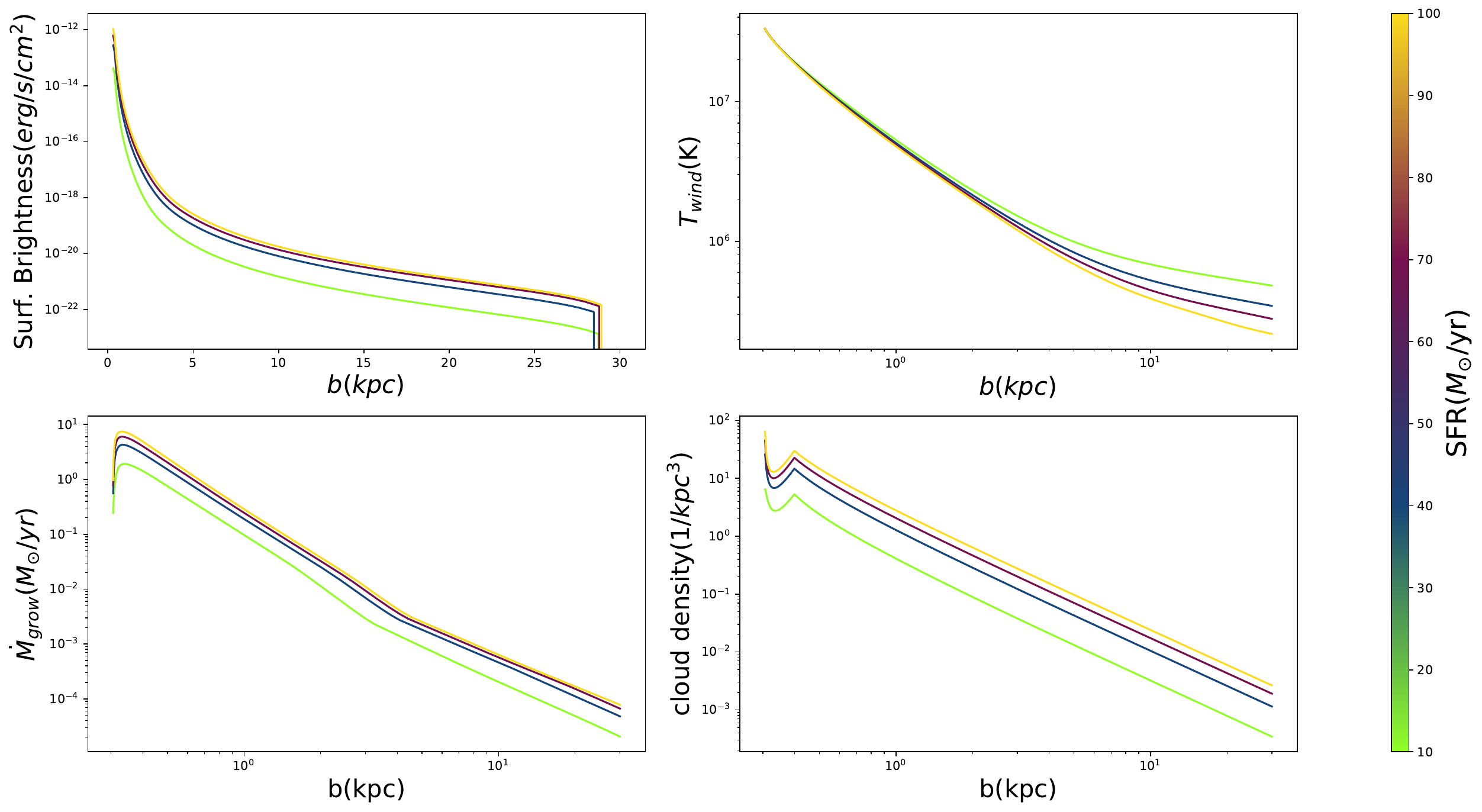}

    \caption{In this panel we present the Surface Brightness along the line of sight for different values of the impact factor $b$, together with different quantities involved in the calculation of the Surface Brightness. The different colors correspond to different Star Formation Rates, while we set $\eta_{\rm mass,hot}=0.3,\eta_{\rm mass,cold}=0.15$. We use for the calculation a a lognormal distribution with $\sigma=1$ and mean initial mass $\rm \sim 10^{5.5} M_{\odot}/yr$. } 
    \label{surface brightness}
\end{figure*}

Figure~\ref{Luminosity1} shows that the luminosity of the halos has values ranging from $\sim 10^{41} \rm erg\;s^{-1}$ to $\sim 10^{42} \rm erg\;s^{-1}$ which agrees well with the observations \citep[e.g.,][]{2010ApJ...717..289S,Wisotzki_2018}. Luminosity increases with the increase of the star formation rate, which is expected as $\dot{M}_{\rm grow}$ also increases with SFR. We also observe that the luminosity decreases for higher values of $\eta_{\rm mass,hot}$. With the initial cold mass flux ratio fixed to $\eta_{\rm mass,cold}=0.15$ this is expected because by increasing $\eta_{\rm mass,hot}$, we get to the regime where we have more hot mass in the beginning, comparing to the cold mass. By having less cold mass spread in the volume, there are less wind-cloud interactions happening, resulting to less cooling, so the luminosity drops. The whole contribution to the luminosity value comes from radii near the initial injection of the clouds. This can be seen by the fast drop on the surface brightness profiles as well, from which we can deduce the reasons behind it.

In Fig.~\ref{surface brightness}, we study the solution for the surface brightness profiles. The calculations done here are presented in physical units, because we want to examine the behaviour of the solution, together with the components involved in the formula. The surface brightness depends on the value of the growth term, and the total number density of the cloud populations. We observe that the values of  $n_{\rm cloud}, \dot{M}_{\rm grow}$ drop several orders of magnitude during the integration interval. From this behaviour, we can correlate the drop in the surface brightness profiles, as well as the contribution from the central bins to the luminosity, to the same drops in the $n_{\rm cloud}, \dot{M}_{\rm grow}$.

By switching to astronomical units and including a red-shift parameter to the surface brightness profiles, we can compare the results with observed \Lya surface brightness profiles \citep{2010ApJ...717..289S,Wisotzki_2018,Lujan_Niemeyer_2022}. However because of the same drop in the surface brightness calculated in our model, the results do not match the observations which typically are much flatter; dropping around only two orders of magnitude in surface brightness levels in the same range of impact parameter (e.g best fit from \citep{Wisotzki_2018}). Thus, we observe a significant difference between the surface brightness profiles of the model and the observation fits. 

In conclusion, while cooling, multiphase galactic winds can produce significant \Lya emission, this is too centrally peaked in order to explain observed \Lya  halos. This strongly suggests that radiative transfer effects are at play leading to the observed \Lya halo shapes \citep[e.g.,][]{Chang_Yang_Seon_Zabludoff_Lee_2023,2023MNRAS.tmp.1720B} -- maybe not unsurprising given the large \Lya optical depth.

\section{Discussion}
In this section, we will analyze here some phenomena that arise with the introduction of multiple cloud populations. We will describe some characteristics and implications of the model, and discuss future directions.

\subsection{Negative relative velocities}
\begin{figure}
\includegraphics[width=0.9\linewidth]{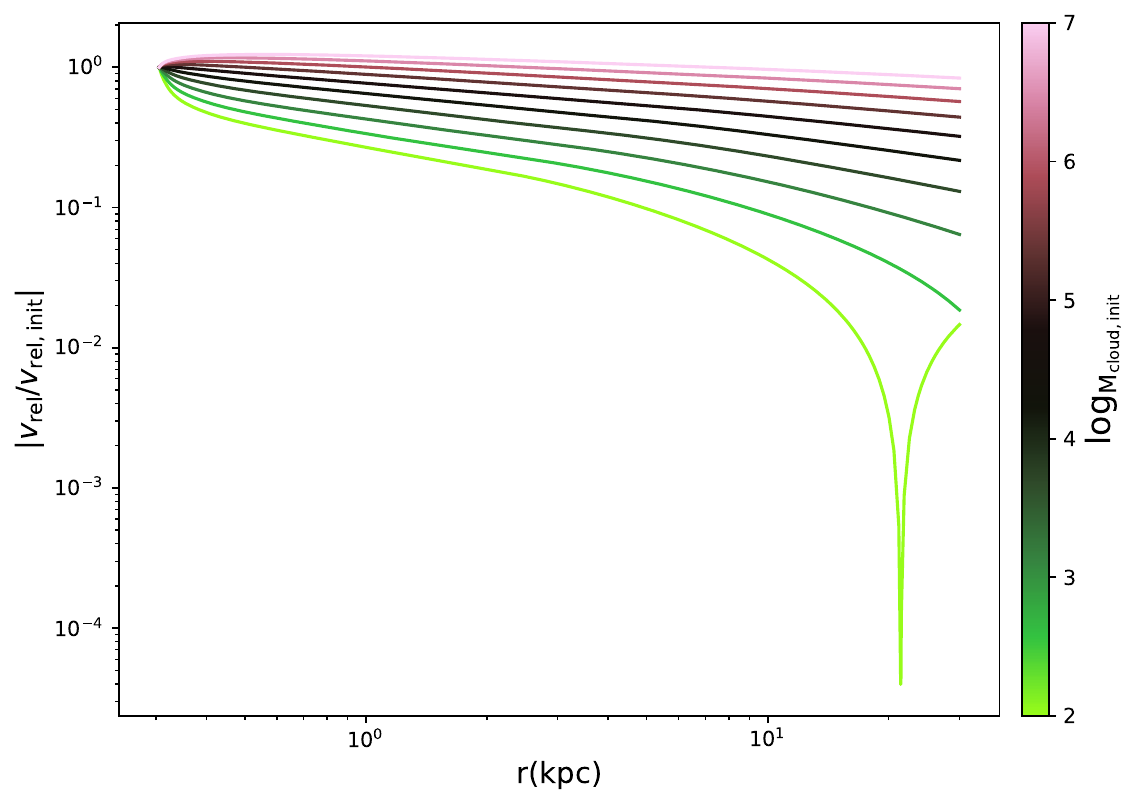}
    \caption{The absolute velocity profile for a SFR=$15M_\odot/\rm yr$,\;$\eta_{\rm mass,hot}=0.4$,\;$\eta_{\rm mass,cold}$=0.1 and a lognormal distribution with range $M_{\rm cloud,init}=10^2-10^7M_\odot$ and $\mu=5,\sigma=1.5$, with the lightest cloud being pushed to negative relative velocities.} 
   \label{fig:relative_vel}
\end{figure}
A new behaviour for the case of multiple populations, is that the heavy clouds influence the lightest clouds. As we can also see in Fig.~\ref{fig:relative_vel}, the lightest cloud populations in the distribution sometimes get accelerated rapidly and they get velocities higher than the wind ones. 

Even though all the equations are automatically zero when $v_{\rm rel}=0$, the behaviour near zero is important in this case. For the cloud velocity to never exceed the wind value, the terms in the cloud velocity equation must get asymptotically to zero as $v_{\rm rel}$ approaches zero. If this is not happening, then the light clouds can get accelerated beyond the wind velocity. In order to explain the phenomenon mathematically, one has to focus on the $\dot{M}_{\rm grow}$ appearing in the cloud velocity equation~\ref{cloud_velocity} due to momentum exchange. 

Because of the backreaction of the clouds on the wind, the gradient of the hot gas velocity can be steeper than the dynamical timescale of a clump population, thus leading to an `overshoot' of the velocity. This happens when $t_{\rm dyn,hot}\sim \ell/\Delta v$ over a certain lengthscale $\ell$ is shorter than the acceleration time of the cloud -- which is $t_{\rm drag}\sim \chi r_{\rm cl}/ v_{\rm rel}$ or $t_{\rm grow}\equiv M_{\rm cloud}/\dot{M}_{\rm grow}$ for ram pressure or momentum transfer, respectively.

Physically, this effect can be easily understood: as the clouds carry more momentum, they have a finite deceleration time -- akin to bullets flying through air.

In order to determine whether this effect is physical or not, it could be needed to analyze the implications of the pulsations for the system. Further work understanding the momentum exchange term of the two phases, might point to the correct direction. The total effect of the phenomenon to the whole system though is not big, because it appears for mass values far from the mean, where the number density of the population undergoing this acceleration is significantly lower.

\subsection{Reasons for cloud destruction}
A general feature of our model is that we observe clouds being destroyed in the whole range of the parameter regime.

The domination of the loss term $\propto v_{\rm rel}$ over the growth term $\propto v_{\rm in}$, is happening mainly because of the saturation of the turbulent velocity $v_{\rm turb}\propto c_{\rm s,h}$, as explained in previous sections. The growth term in that sense, gets a rather complicated form with different cappings. On the other hand, the loss term follows the simple form from linear theory, with the extension from \citet{Br_ggen_2016}. A point forward would be to investigate more the loss term through simulations, especially for high overdensities $\chi\gtrsim 1000$, in order to understand the nature of the loss term better. 

In the same sense, a limitation of our work that might lead to the cloud destruction is the use of both $\dot{M}_{\rm grow}$ and $\dot{M}_{\rm loss}$ simultaneously. More precisely, both growth and loss equations have been tested in cases where growth or destruction dominates, respectively, but the presence of both of them together is an assumption. For example the growth equation is derived from an application of the Gauss law, and one has to assume a one way flow of mass between the two components. Another approach for the model could be used, with each term dominating when there is survival or destruction only, but then the decision of which term dominates and when, is arbitrary. A model that could lead to these approximations physically is more ideal but far more challenging to construct.

\subsection{Connection to observations}

\begin{figure}
\includegraphics[width=0.9\linewidth]{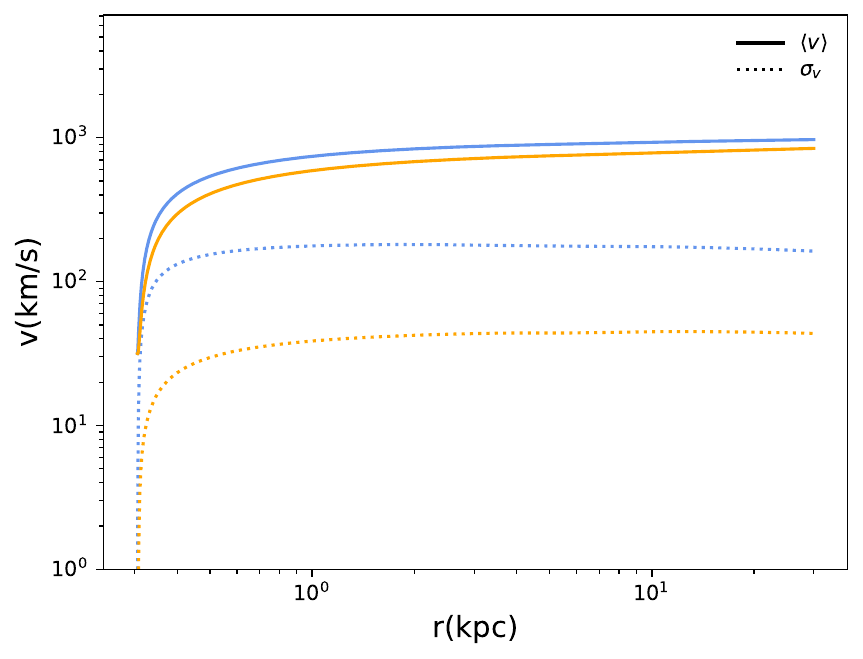}
    \caption{We show the mean velocity (straight lines) and the root mean square velocity (dashed lines). The quantities are presented for two different distributions: a lognormal distribution (blue) and a powerlaw distribution (orange). We use SFR=$15M_\odot/\rm yr,\;\eta_{\rm mass,hot}=0.4,\;\eta_{\rm mass,cold}=0.1$. We choose a powerlaw $\propto M_{\rm cloud}^{-2}$. The lognormal distribution has an initial mass range of $M_{\rm cloud,init}=10^2-10^7M_\odot$ and $\sigma=1.0$. Both distributions have initially same mean mass $\sim 10^{5.5} M_\odot$} 
   \label{dispersion}
\end{figure}

In section~\ref{subsection:observations}, we computed the surface brightness profile originating from cooling radiation because of the mixing and inflow of hot gas. The vast majority ($\gtrsim 90\%$) of this energy will be radiated away via Lyman-$\alpha$ emission of neutral hydrogen.

As stated, we generally find that while the overall energy budget is non-negligible ($\sim 10^{42}\,{\rm erg}\,{\rm s}^{-1}$), the surface brightness dropoff is much steeper than observed in high-$z$ \Lya halos \citep{Steidel2011,Wisotzki_2018}. As we explore in section~\ref{subsection:observations}, this is due to the fact that both the temperature as well as the $\dot{M}_{\rm grow}$ drop (by several orders of magnitude; cf. Fig.\ref{surface brightness}).

For other emission lines, the fraction of cooling radiation is of course even lower. This implies that extended H$\alpha$ emission observed in (local) galaxies are predominantly powered by recombination -- as suggested by previous studies \citep[e.g.][]{1987AJ.....93..264M,Strickland_Heckman_Weaver_Hoopes_Dahlem_2002,Hoopes_Heckman_Strickland_Howk_2003}. For instance, the H$\alpha$ surface brightness levels of M82 are between $10^{-14}\sbu$ and $10^{-16}\sbu$ at distances up to $\sim 5\,$kpc from the nucleus \citep{Yoshida_Kawabata_Ohyama_Itoh_Hattori_2019}. For this redshift ($z\sim 0$) we find total surface brightness levels of $10^{-14}\sbu$ to $10^{-18}-10^{-19}\sbu$, i.e, 2 and 3 orders of magnitude lower. Note that the H$\alpha$ surface brightness will be approximately another two orders of magnitudes lower than the total computed here \citep[e.g.][]{2011piim.book.....D,Bustard_Gronke_2022}.
Whether these recombination events are due to photoionization or shocks is beyond this study but discussed extensively in the literature \citep[e.g.][]{1987AJ.....93..264M}.

Nevertheless, emission studies of galactic winds can help constrain our model further, e.g., by nailing down the clump size distribution. Here, we agnostically studied both lognormal as well as powerlaw (with $-2$ slope) distributions with the latter being supported by theory \citep{Gronke_Oh_Ji_Norman_2021,2023arXiv230514424T} as well as commonly observed in the ISM \citep[e.g.][]{1955ApJ...121..161S,2010ApJ...710.1247S}. However, what the clump size distribution is in winds is still unclear but can be constrained with current and future observations of local winds.

Another way of observing outflows, is naturally through `down the barrel' absorption \citep[e.g.,][]{2005ApJ...621..227M,2015ApJ...811..149C,Li2023}. Here,  the comparison of theory and observations crucially relies on the mapping between real and velocity space and thus on models such as the one presented here. One particular interesting fact of the multi-cloud model is the increased apparent `velocity dispersion' due to the differential acceleration. That is, because differently sized clouds accelerate at different rates, at a fixed distance (and thus also as an integrated measure), there is not just a fixed (mean) velocity but also a potentially large dispersion around that mean.

We illustrate this in Fig.~\ref{dispersion}, where we show the mean as well as the `root mean square' velocity of one particular model as a function of radius. Clearly, the variance of the velocities is non negligible -- and crucially depends on the cloud distribution. We can see an order of magnitude difference between a log-normal (blue lines) and a powerlaw (orange lines) -- in spite of the fact that their mean velocities (shown as solid lines in Fig.~\ref{dispersion}) are comparable. Interestingly this dispersion is rather large ($\sim 100 \Kms$) and, thus, potentially larger than usual turbulence in galactic winds \citep[e.g.,][found values of $\lesssim 30\Kms$]{2020ApJ...895...43S}. A similar value for turbulent winds ($\sim 100\Kms$) was found by \citet{2023arXiv230514424T}. Of course this dispersion is not isotropic and thus not a turbulent component, however, this is impossible to disentangle from down-the-barrel observations alone. 

In conclusion, our model can provide insights into observations by directly comparing (mock) observations to theoretical expectations. This is specifically useful for absorption line studies where our multi-cloud model can explain potentially large velocity dispersions required to explain observations \citep[see, e.g.,][]{Li2023}.

\subsection{Caveats of the model}
Semi-analytic models such as ours crucially rely on basic theory and simulation work to map out the viable parameter space and provide the scalings used. It is, therefore, also important to cross-check the model predictions with results of these previous studies -- something we tried to achieve in section~\ref{subsection:homogeneous} by comparing a homogeneous background, single-cloud case of our model to `cloud crushing' simulations.

However, this has not been done in the entire (realistic) parameter space for a lack of such basic studies. In particular,
\begin{itemize}
    \item most simulation work focus on the low $\chi\sim 100$, low $\mathcal{M}$ regime \citep[e.g.,][]{1993ApJ...407..588M,2000ApJ...543..775G,Abruzzo_2022}. However, the winds of our model usually show much larger overdensities ($\chi\gtrsim 1000$) and Mach numbers (throughout the evolution $\mathcal{M}\sim 3-10$). More simulation is needed there to calibrate semi-analytical models in this regime.
    
    \item similarly, most cloud-wind simulations focus on a homogeneous background (exceptions include \citealp{Gronke_2019} for outflowing and \citealp{Heitsch_Putman_2009,Gronnow2017,Heitsch2021,Tan_Oh_Gronke_2023} for infalling clouds). However, clearly a realistic wind background is varying and more work is needed in this direction.

    \item while arguably the $v_{\rm in}$ term is well understood from detailed simulations of turbulent mixing layers \citep{Fielding_2020,Tan_2021}, the (evolution of the) surface area of the cold gas is much less so. Our model includes a temporal variation of the surface area, however, this is clearly a simplification of the fragmentation and coagulation processes \citep{2022arXiv220900732G} occurring in multiphase systems. This is in particular true for clouds close to the destruction limit, i.e., where neither the growth nor the destruction term can be neglected -- making essentially all of our clouds hard to model.

    \item a related point is the neglection of `ensemble effects', that is, the influence of clouds (and cloud populations) directly upon each other. Our model includes the indirect (i.e., via the wind) influence of cloud populations but not the (arguably more important; \citealp{Cowie1981,Aluzas2012,Poludnenko2002,Banda2020,2023arXiv230514424T}) direct ones. This clearly requires more (simplified) simulation work to study these effects.

    \item other physical processes we neglected here are magnetic fields, cosmic rays, viscosity and thermal conduction. While, e.g., viscosity and thermal conduction might not affect directly $v_{\rm in}$ \citep[see][for the inclusion of thermal conduction in turbulent mixing layer simulations]{Tan_2021}, they can alter the surface area drastically and, thus, $\dot{M}_{\mathrm{grow}}$ \citep{Bruggen_Scannapieco_Grete_2023,Br_ggen_2016,Scannapieco_2015}. Similarly, magnetic fields can play a large role in cold gas survival \citep{McCourt_OLeary_Madigan_Quataert_2015,Shin_Stone_Snyder_2008,2023arXiv230409897H}.

\end{itemize}

While our model provides a next step towards realism, and is useful to compare theory to observations, it is important to keep these caveats in mind. Overcoming them will require more theoretical and computational efforts of the community.

\section{Conclusions}
We present an analytical model for the interaction of the wind with multiple cloud populations extending previous work by \citetalias{Fielding_2022}. In order to model the interactions between the phases as well as the time dependent area of the clouds, we used findings from small scale hydrodynamic simulations. Specifically, we implemented scalings found in high-resolution turbulent mixing layer simulations by \citet{Tan_2021} and compared the final model to `cloud crushing' simulations \citep{Gronke_2018}. Furthermore, we introduced probability distributions to describe the number density of each cloud population. Finally, we made an attempt to connect our model with observations, by calculating the emitted surface brightness profiles.

Our main findings are:
\begin{itemize}
    \item In the homogeneous case we can reproduce results from hydrodynamic simulations fairly well in the $\chi \sim 100$ and with  slight deviations in the high $\chi\sim1000$ case. For the whole parameter range of the initial conditions, the $t_{\rm cool,mix}/t_{\rm cc}$ survival criterion is satisfied except for high $\chi$, $\mathcal{M}$ in the region between ${t_{\rm cool,mix}}/{t_{\rm cc}}\in [0,10]$.
    \item For the single cloud population in an adiabatically expanding hot wind background, clouds have the tendency to die, because $\dot{M}_{\rm loss}\propto v_{\rm rel}$ dominates.  For high SFR and $\eta_{\rm M,hot}$ lighter clouds at first accelerate rapidly and gain mass but always start losing mass after some point in the evolution. Heavier clouds have a more steady evolution for the whole parameter range.
    \item The main difference compared to the  \citetalias{Fielding_2022} model, when we use a single cloud population, is the saturation of the inflow velocity $v_{\rm in}$, together with a dynamical evolution for the length of the tail.
    \item For the multiple cloud case, the clouds have again a tendency to get destroyed. The choice of the distribution is important for certain values of the initial conditions, mainly the ones that we see significant differences in the evolution of each cloud population. For higher values of SFR and $\eta_{\rm mass,hot}$ the differences are not that significant and the model is not very sensitive to the choice of cloud distribution. Curiously, in some cases the lightest clouds are ``pushed'' to negative relative velocities, mainly when there is a big range of initial cloud masses, that span several orders of magnitude.
    \item The cooling luminosity values are around $10^{42}\rm erg\;s^{-1}$, while almost all of the contribution to this value comes from the central region. The reason for that is the drop in the $\dot{M}_{\rm grow}$ and $n_{\rm cloud}$ term, after the first few kpc. The same effect can be seen in the surface brightness profiles. Due to this, there is a significant difference between emissions due to cooling and the observed \Lya halos as well as measured H$\alpha$ profiles indicating radiative transfer and photionization effects (cf. \S~\ref{subsection:observations}).
\end{itemize}
Our analytic multiphase galactic wind model captures the variety of cloud sizes a realistic galactic winds do possess. However, our work also shows that more work is needed in particular in constraining the source and sink terms via simplified, multiphase simulations.

\section*{Acknowledgements}
The authors acknowledge support through the MPA internship program, which made this research possible. The authors thank Chad Bustard, Brent Tan, Maja Lujan Niemeyer and Seok-Jun Chang for useful discussions. The authors also thank the anonymous referee for useful comments that improved the manuscript. MG thanks the Max Planck Society for support through the Max Planck Research Group. CN acknowledges support from DAAD, through the scholarship programme ``Master studies for all academic disciplines''. This research was supported in part by grant NSF PHY-2309135 to the Kavli Institute for Theoretical Physics (KITP). This research has made use of the publicly available code provided by \citetalias{Fielding_2022}, as well as Matplotlib: a Python library for publication quality graphics \citep{Hunter:2007}, NumPy \citep{harris2020array}, SciPy \citep{2020SciPy-NMeth} and the colormaps in the CMasher package \citep{2020JOSS....5.2004V}.

\section*{Data Availability}
Data and code related to the results of this work will be shared on reasonable request to the corresponding author.



\bibliographystyle{mnras}
\bibliography{example} 




\appendix
\section{Survival Criterion for different radii}
\label{app:criterion}
\begin{figure*}
\includegraphics[width=0.9\linewidth]{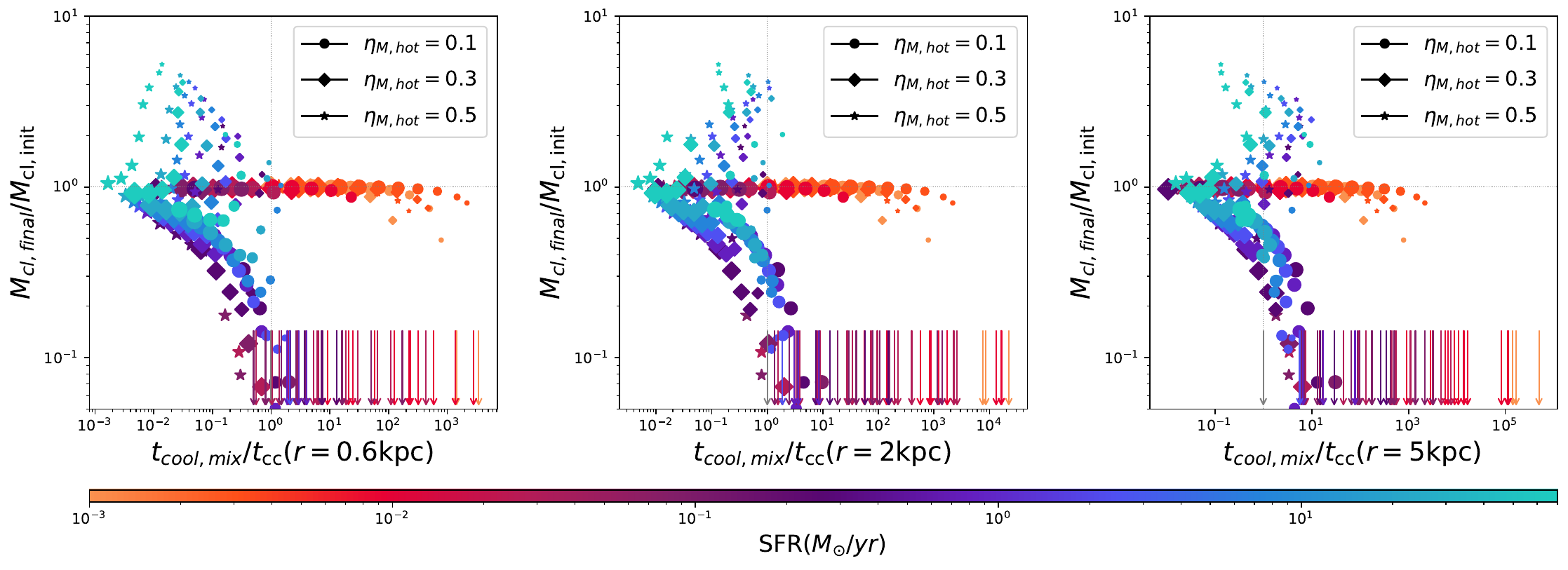}
    \caption{In this panel, we present the model solutions for a cloud mass evaluated at $r\sim 30 \mathrm{kpc}$ and different values of $t_{\rm cool,mix}/t_{\rm cc}$. In each plot, $t_{\rm cool,mix}/t_{\rm cc}$ is calculated at different radius: $r=0.6,2,5\mathrm{kpc}$ respectively.The initial cloud masses range from $10^0.5(M_\odot)$ to $10^7(M_\odot)$ and every marker's size is proportional to the corresponding mass. The Star Formation Rate ranges from SFR=$0.001(M_\odot/\rm yr)$ to $70(M_\odot/\rm yr)$ and the $\eta_{\rm mass,hot}=[0.1,0.3,0.5]$ }
   \label{scatter-panel}
\end{figure*}
In Figure~\ref{fig:scatter-single}, the survival criterion of \citet{Gronke_2018} seems to not hold in the dynamical wind case. The value of the $t_{\rm cool,mix}/t_{\rm cc}$ in that case is calculated at the injection point for the clouds ($r\sim300\mathrm{pc}$). We proceed to check the behaviour of the criterion, if we calculate $t_{\rm cool,mix}/t_{\rm cc}$, at larger radii, where the density of the clouds falls, and the conditions become more idealized.

In Figure~\ref{scatter-panel}, we calculate the cloud mass $M_{\rm cl,final}(r=30\mathrm{kpc})/M_{\rm cl,init}(r=0.3\mathrm{kpc})$ and present it versus the $t_{\rm cool,mix}/t_{\rm cc}$ evaluated at three different radii: ($r=0.6\mathrm{kpc},r=2\mathrm{kpc}$, and $r=5\mathrm{kpc}$). Because the model has a general destructive behavior, even the points with $M_{\rm cl,final}/M_{\rm cl,init}>1$ will eventually die. Thus, we would expect most values to be in the down and right quarter. By moving to larger values of the radius, the points are shifted to the right. We can see that the agreement becomes better for larger radii albeit with a large scatter. 

\section{Different Overdensities}
\label{app:different_chis}
We compare the solutions of our model in the homogeneous case, with different simulations carried out by \citet{Gronke_2018}, to show explicitly the disagreement that emerges for higher values of the overdensity $\chi$.
\begin{figure*}
	\includegraphics[width=0.7\linewidth]{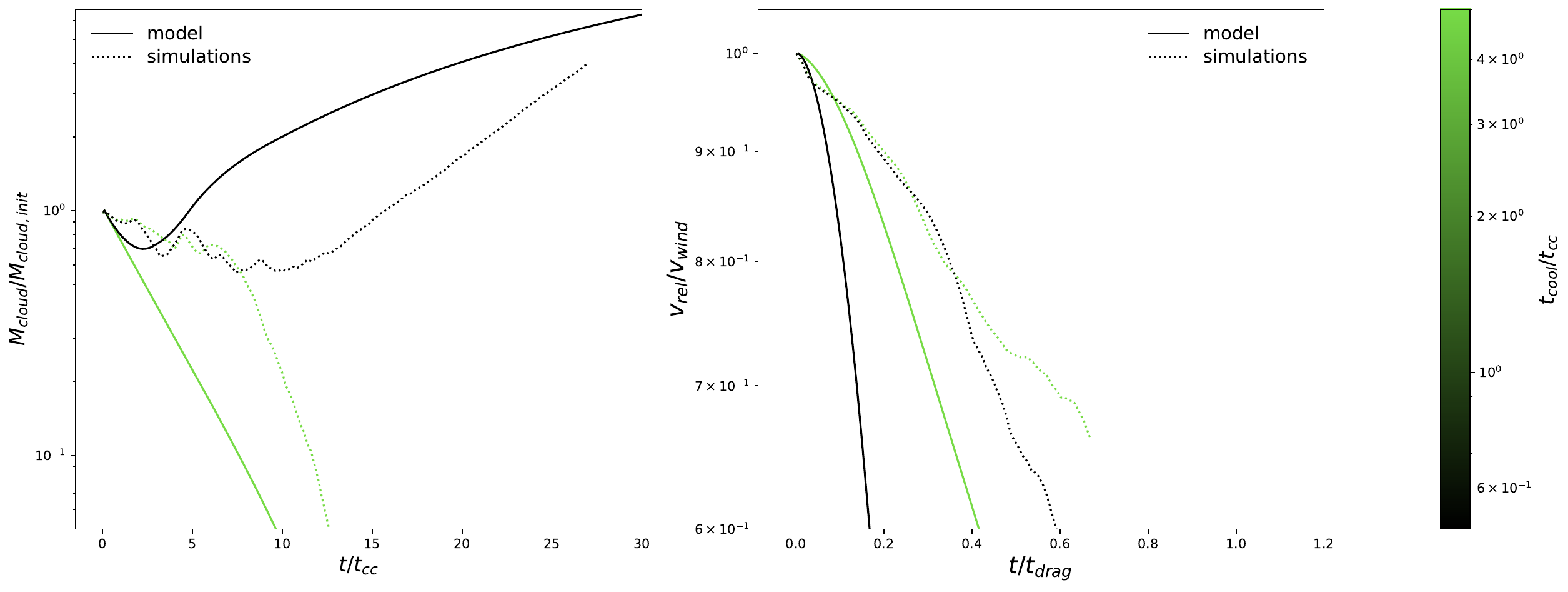}
    \caption{The time evolution of the cloud mass is presented in the left hand panel, while the time evolution of the relative velocity is presented in the right hand panel. The solutions from the model presented in this paper are presented in straight lines in comparison with the results of hydrodynamic simulations from \citet{Gronke_2018} which are presented in dashed. In order to use the same conditions as the hydrodynamic simulations, we use $\chi=300, T_{\rm cloud}=4\cdot10^4\text{K}$ and  wind Mach number $\mathcal{M}=1.5$. The Pressure of the wind is $P_{\rm wind}/\text{kb}=10^4 \text{K}/\text{cm}^3$. Different initial values for $t_{\rm cool,mix}/t_{\rm cc}$ are presented with different colors.} 
    \label{fig:comparison300}
    \end{figure*}
    \begin{figure*}
	\includegraphics[width=0.7\linewidth]{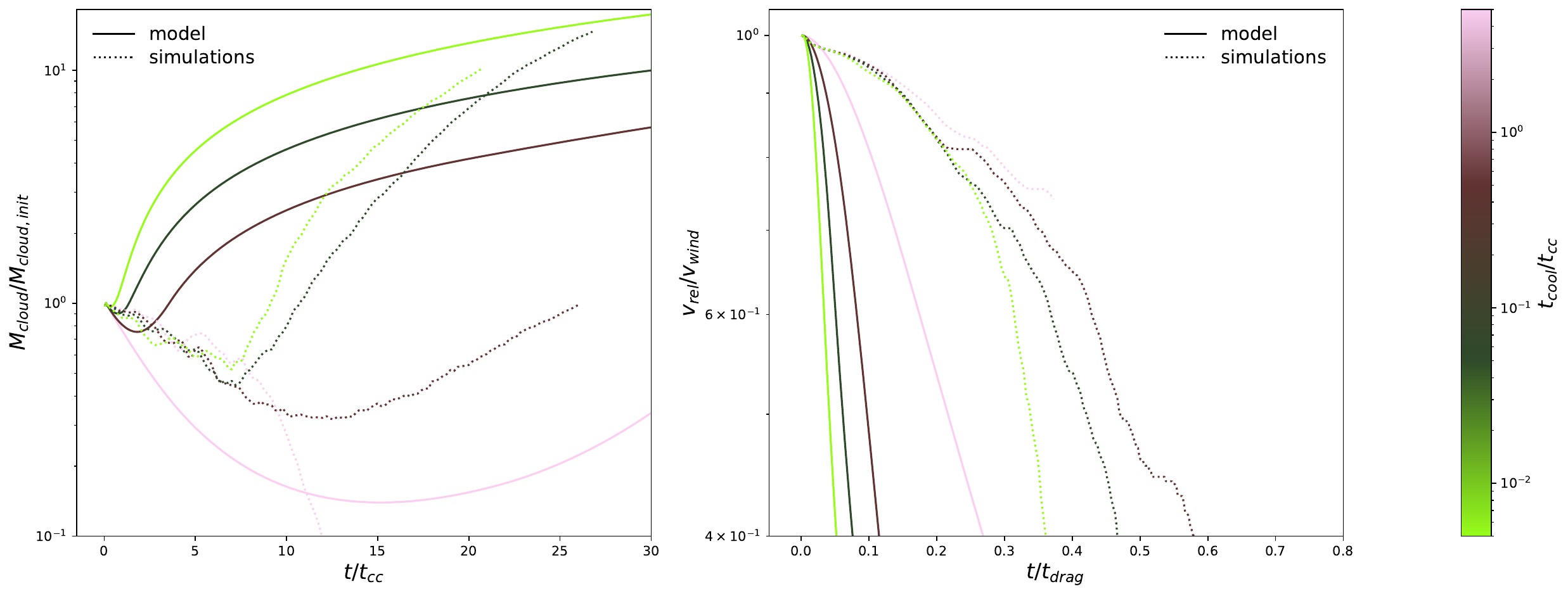}
    \caption{The same evolution panel for the cloud mass (left panel) and the relative velocity (right panel) as in Fig.~\ref{fig:comparison300}, for a different overdensity value, $\chi=1000$. The solutions from the model presented in this paper are presented in straight lines in comparison with the results of hydrodynamic simulations from \citet{Gronke_2018} which are presented in dashed. Different initial values for $t_{\rm cool,mix}/t_{\rm cc}$ are presented with different colors.} 
    \label{fig:comparison1000}
\end{figure*}
Figure~\ref{fig:comparison300}, shows the cloud mass and relative velocity evolution for two different values of $t_{\rm cool,mix}/t_{\rm cc}=0.5$  and $t_{\rm cool,mix}/t_{\rm cc}=5$, for overdensity $\chi=300$ and $\mathcal{M}=1.5$. We see that in this case, even though there are differences in the form of the cloud mass function, our model manages to  reproduce the general behaviour of the simulations: Clouds grow for the $t_{\rm cool,mix}/t_{\rm cc}<1$ and die otherwise.

Figure~\ref{fig:comparison1000} exhibits the case of overdensity $\chi=1000$. Here four different values of the ratio are presented with different colours: $t_{\rm cool,mix}/t_{\rm cc}=0.005,0.05,0.5,5$. In this high $\chi$ case, we can see a slight disagreement of our model with simulations. More precisely, for the $t_{\rm cool,mix}/t_{\rm cc}=5$ case, our solution at first loses mass, but eventually regrows, while the simulated cloud loses mass and dies. Interestingly, the solution's behavior is similar to the $t_{\rm cool,mix}/t_{\rm cc}=0.5$ case. This disagreement is the same as the one discussed in Section~\ref{subsection:homogeneous}, where in Figure~\ref{fig:scatter-uniform}, we see a discrepancy of our solutions and the criterion over the region $t_{\rm cool,mix}/t_{\rm cc}\in [1,10]$. Apart from that case, we see that the rest of the solutions follow the behavior of the simulations and clouds grow, although the precise evolution of the clouds is different from \citet{Gronke_2018}.


\bsp	
\label{lastpage}
\end{document}